\documentclass[aps,pra,reprint,superscriptaddress,showpacs,uplatex]{revtex4-1}

\usepackage{amsmath,amssymb,bm}
\usepackage{graphicx}

\begin{document}

\title{Decomposition of radiation energy into work and heat}

\author{Tatsuro Yuge}
\email[]{yuge.tatsuro@shizuoka.ac.jp}
\affiliation{Department of Physics, Shizuoka University, Suruga, Shizuoka 422-8529, Japan}

\author{Makoto Yamaguchi}
\affiliation{Center for Emergent Matter Science, RIKEN, Wako, Saitama 351-0198, Japan}

\author{Tetsuo Ogawa}
\affiliation{Department of Physics, Osaka University, Machikaneyama, Toyonaka, 560-0043, Japan}

\date{\today}

\begin{abstract}
We investigate energy transfer by the radiation from a cavity quantum electrodynamics (QED) system
in the context of quantum thermodynamics.
We propose a method of decomposing it into work and heat within the framework of quantum master equations.
We find that the work and heat correspond respectively to the coherent and incoherent parts of the radiation.
In the derivation of the method,
it is crucial to investigate the dynamics of the system that receives the radiation from the cavity.
\end{abstract}


%03.65.-w	Quantum mechanics
%03.65.Yz	Decoherence; open systems; quantum statistical methods
%
%05.30.-d	Quantum statistical mechanics
%05.70.-a	Thermodynamics
%05.70.Ln	Nonequilibrium and irreversible thermodynamics
%05.90.+m	Other topics in statistical physics, thermodynamics, and nonlinear dynamical systems (restricted to new topics in section 05)
%
%07.20.Pe	Heat engines; heat pumps; heat pipes
%
%42.50.-p	Quantum optics
%42.50.Lc	Quantum fluctuations, quantum noise, and quantum jumps
%42.50.Pq	Cavity quantum electrodynamics; micromasers
%42.55.Ah	General laser theory
%42.60.Lh	Efficiency, stability, gain, and other operational parameters
%
%88.05.Bc	Energy efficiency; definitions and standards
%88.05.De	Thermodynamic constraints on energy production

\maketitle

\section{Introduction}

Thermodynamics is an established and powerful phenomenological theory in the macroscopic scale
\cite{Fermi, Callen, GyftopoulosBeretta}.
It provides various universal results on thermodynamic processes:
e.g., the Carnot efficiency limit of heat engines.

In the 19th Century, one of the key steps in constructing thermodynamics
was to realize the equivalence and difference between work and heat \cite{Clausius, Thomson}.
The equivalence---both are forms of energy transfer---led to the first law,
and the difference---100\% conversion from heat to work is impossible---led to the second law and furthermore to the concept of entropy.

After its establishment as the reliable theory on macroscopic physics,
thermodynamics has played many important roles in the relationship to quantum physics.
They range from the one in the genesis of quantum physics \cite{Einstein}
to those in the currently emerging field of quantum thermodynamics \cite{Kosloff, Mahler, VinjanampathyAnders, Goold_etal}.
A perspective on the relationship is
how to define and incorporate work, heat, and entropy in quantum physics.
In an early stage, entropy in quantum systems was introduced by von Neumann \cite{vonNeumann}
through a thermodynamic consideration.
An example of the utilizations of entropy in quantum systems
was seen in 1970s in discussing stability of equilibrium states \cite{Araki, HatsopoulosGyftopoulos}.
Stability was also studied in terms of work done by quantum systems, which led to the concept of passivity \cite{PuszWoronowicz, Lenard}.
Around the same time, related studies in open quantum systems appeared.
There, entropy balance \cite{Alicki1976, Spohn} and its relation to
energy balance (work--heat decomposition) \cite{SpohnLebowitz, Alicki1979, Kosloff1984} were investigated.
In particular, Refs.~\cite{Alicki1979} and \cite{Kosloff1984} studied heat engines in the setup of open quantum systems.
Before these studies, a quantum heat engine was studied \cite{ScovilSchulz-DuBois, Geusic_etal}
in the context of population inversion or negative temperature \cite{Rmasey}.

Now in the 21th Century, renewed attention has been paid to quantum thermodynamics
\cite{Kosloff, Mahler, VinjanampathyAnders, Goold_etal,
AllahverdyanNieuwenhuizen, Gemmer_etal, Horodecki_etal, Scully_etal2003, Allahverdyan_etal, Weimer_etal,
Scully, Scully_etal2011, Beretta, Harbola_etal, Dorfman_etal, Gelbwaser-Klimovsky_etal, GoswamiHarbola,
Aberg, Correa_etal, Rosnagel_etal, Mitchison_etal, BraskBrunner, Uzdin_etal, Korzekw_etal, Dag_etal}.
One of the main goals of quantum thermodynamics is the extension of thermodynamics
under quantum effects, such as quantum coherence and entanglement.
For example, there are reports that quantum coherence enhances the performance of quantum thermal machines
(such as quantum heat engines and refrigerators)
\cite{Scully_etal2003, Scully, Scully_etal2011, Harbola_etal, Dorfman_etal, Aberg, Correa_etal,
Rosnagel_etal, Mitchison_etal, BraskBrunner, Korzekw_etal, Dag_etal}.

\begin{figure}[t]
\begin{center}
\includegraphics[width=\linewidth]{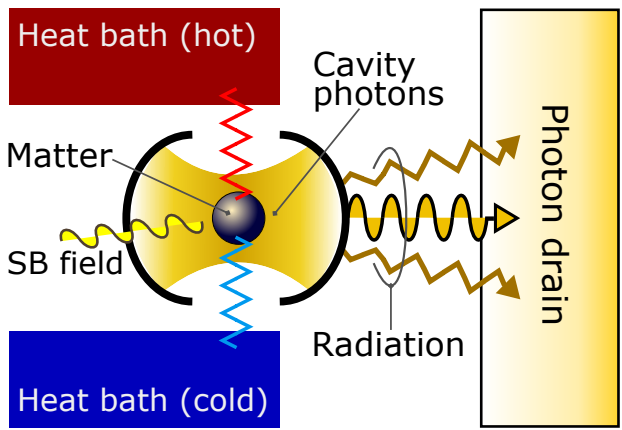}
\end{center}
\caption{(Color online) A typical example of the total system setup.
The cavity QED system consists of the cavity photons and matter.
Both coherent and incoherent light is included in the radiation from the cavity to the photon drain.
The symmetry breaking (SB) field is introduced to keep photonic amplitude finite (see the text).
}
\label{fig:setup}
\end{figure}

Lasers and masers are regarded as examples of quantum heat engines and have been studied
in the context of the quantum thermodynamics
\cite{Scully_etal2011, Dorfman_etal, ScovilSchulz-DuBois, Dag_etal, GoswamiHarbola}.
A typical setup of a laser system as a heat engine is shown in Fig.~\ref{fig:setup}.
The main system is a cavity quantum electrodynamics system, which consists of cavity photons and matter.
The matter is thermally pumped by heat baths.
The output of the engine is obtained in a form of the radiation from the cavity.
Below a certain laser threshold, incoherent light is mainly observed in the radiation.
Above the threshold, by contrast, coherent light is mainly observed.
In this way, the system emits both coherent light (such as a laser) and incoherent light
(such as thermal radiation).

It is noted that the energy transfer by coherent radiation is ``systematic''
whereas that by incoherent one is ``random.''
This leads to an intuitive conjecture that the coherent and incoherent parts
of the radiation correspond respectively to work and heat.
However, so far in many studies on lasers as quantum heat engines,
all the energy transfer by the radiation has been calculated as work
\cite{Scully_etal2011, Dorfman_etal, ScovilSchulz-DuBois}
(an exception is Ref.~\cite{GoswamiHarbola}).
One should carefully define the work and heat in the radiation (output of the engine)
when discussing quantum thermodynamics.
Otherwise, for example, one sometimes overestimates the efficiency of a heat engine.
Decomposing the radiation energy into work and heat should be particularly crucial at around the threshold
because coherent and incoherent parts equally contribute to the radiation.

In the present paper, we propose an appropriate method of decomposing the radiation energy
into work and heat in the framework of quantum master equation (QME).
We formulate the method by investigating the time evolution of the system (called photon drain)
that receives the radiation from the cavity.
We definitely identify the systematic (work) and random (heat) energy transfers based on the time evolution.
(For a similar but different approach, see Ref.~\cite{Gelbwaser-Klimovsky_etal}.)
Our main result shows that the above intuitive conjecture is correct:
the coherent radiation is attributed to the work whereas the incoherent one is attributed to the heat.

\section{Setup}\label{sec:Setup}

We consider a total system which is composed of a cavity quantum electrodynamics (QED) system, heat baths and a photon drain.
In the present paper, we refer to cavity quantum electrodynamics system, heat bath(s), and photon drain simply as cQED, bath(s), and drain, respectively.
We show a schematic example of the setup in Fig.~\ref{fig:setup}.
For simplicity, we assume that the cavity is single mode.
The baths interact with the matter part of the cQED and induce excitation and decay of the cQED.
The drain, on the other hand, interacts with the photon part of the cQED
and receives the photons out of the cQED through the cavity loss.
We also irradiate weak coherent field, which we call symmetry breaking (SB) field, to the cavity photon mode
to have a non-vanishing expectation value of photonic amplitude in lasing situations.
In the present paper we assume that the baths and drain are so large that their states are almost unchanged.
In particular, the drain is always in the (almost) vacuum state.
We consider the steady state of the cQED in this setup.
That is, we investigate the cQED as a continuous heat engine \cite{Uzdin_etal}.

The total Hamiltonian is given by
\begin{align}
\hat{H}_{\rm tot}(t) = \hat{H}_{\rm c} + \hat{H}_{\rm SB}(t) + \hat{H}_{\rm b} + \hat{H}_{\rm d}
+ \hat{H}_\text{c-b} + \hat{H}_\text{c-d}.
\end{align}
Here, $\hat{H}_{\rm c}$ is the Hamiltonian of the cQED.
Its concrete form is irrelevant to our formulation.
$\hat{H}_{\rm SB}$ is the interaction Hamiltonian between the cavity mode and SB field:
\begin{align}
\hat{H}_{\rm SB}(t) = f_{\rm SB} e^{-i\omega_{\ell} t} \hat{c}^\dag + f_{\rm SB}^* e^{i\omega_{\ell} t} \hat{c},
\end{align}
where $\hat{c}$ and $\hat{c}^\dag$ are respectively the annihilation and creation operators of the cavity mode,
$f_{\rm SB} $ is the amplitude of the SB field, and $\omega_{\ell}$ is the frequency of the SB field.
We set $\omega_{\ell}$ equal to the frequency of cQED photonic amplitude in the rest frame.
$\hat{H}_{\rm b}$ is the free Hamiltonian of the baths and $\hat{H}_\text{c-b}$ is the interaction Hamiltonian between the (matter part of) cQED and baths.
$\hat{H}_{\rm d}$ is the free Hamiltonian of the drain:
\begin{align}
\hat{H}_{\rm d} = \sum_k \hbar \omega_k \hat{d}_k^\dag \hat{d}_k,
\end{align}
where $\hat{d}_k$ and $\hat{d}_k^\dag$ are respectively the annihilation and creation operators of the $k$th mode in the drain.
$\hat{H}_\text{c-d}$ is the interaction Hamiltonian between the cQED and drain:
\begin{align}
\hat{H}_\text{c-d} = \sum_k (\hbar g_k \hat{c}^\dag \hat{d}_k + \hbar g_k^* \hat{c} \hat{d}_k^\dag),
\end{align}
where $g_k$ represents the coupling strength.
We also define the spectral function of the drain:
\begin{align}
\Gamma_{\rm d}(\omega) \equiv 2\pi \sum_k | g_k |^2 \delta(\omega - \omega_k).
\label{spectral_drain}
\end{align}

As seen in the above, the explicit forms of the Hamiltonians are specified only for $\hat{H}_{\rm SB}$, $\hat{H}_{\rm d}$, and $\hat{H}_\text{c-d}$.
The forms of the other Hamiltonians are irrelevant to the present study
although there are some (weak) restrictions on them.
One of the restrictions is that the Hamiltonians other than $\hat{H}_{\rm SB}$ commute with the excitation number of the total system $\hat{N}_{\rm tot}$:
\begin{align}
[\hat{H}_j, \hat{N}_{\rm tot}] = 0,
\label{excitation_conservation}
\end{align}
where $j=$ c, b, d, c-b, and c-d.
Note that $\hat{N}_{\rm tot}$ is defined by the sum of the excitation numbers in the individual systems:
$\hat{N}_{\rm tot} = \hat{N}_{\rm c} + \hat{N}_{\rm b} + \hat{N}_{\rm d}$,
where $\hat{N}_{\rm d} \equiv \sum_k \hat{d}_k^\dag \hat{d}_k$
and $\hat{N}_{\rm c} \equiv \hat{c}^\dag \hat{c} + \hat{N}_{\rm matter}^{\rm ex}$
($\hat{N}_{\rm matter}^{\rm ex}$ is the excitation number of the matter part in the cQED).

We here mention the reason to introduce the SB field.
If we analyze the setup without the SB field,
the photonic amplitude (expectation value of $\hat{c}$) vanishes in the steady state even in lasing situations
because the system has U(1) symmetry (excitation number conservation) as assumed in Eq.~(\ref{excitation_conservation}).
However, as we will see later, the photonic amplitude plays a crucial role in our results in the present paper.
For this reason, finite value of the amplitude is necessary in our formulation,
and this is why we introduce the SB field.

For the convenience of the theoretical analysis,
we transform the frame from the rest one to the rotating one with the frequency $\omega_{\ell}$.
We define an operator $\tilde{A}$ in the rotating frame by
$\tilde{A}(t) = e^{i \omega_{\ell}t \hat{N}_{\rm tot}} \hat{A}(t) e^{-i \omega_{\ell}t \hat{N}_{\rm tot}}$
for any operator $\hat{A}$ in the rest frame.
In this transformation, we have a time-independent total Hamiltonian:
$\tilde{H}_{\rm tot} = \tilde{H}_{\rm c} + \tilde{H}_{\rm SB} + \tilde{H}_{\rm b} + \tilde{H}_{\rm d}
+ \tilde{H}_\text{c-b} + \tilde{H}_\text{c-d}$.
Here,
\begin{align}
\tilde{H}_{\rm SB} = f_{\rm SB} \hat{c}^\dag + f_{\rm SB}^*  \hat{c}.
\end{align}
The other Hamiltonians are unchanged in this transformation because they commute with $\hat{N}_{\rm tot}$
as assumed in Eq.~(\ref{excitation_conservation}).
In the following, we describe the system in the rotating frame unless otherwise specified,
and omit the tildes for notational simplicity.
We also employ the Schr\"odinger picture.

\section{Quantum master equation for cQED}

The time-evolution equation for the total system is given by the von Neumann equation:
\begin{align}
\frac{d}{dt} \rho_{\rm tot}(t)
= \frac{1}{i\hbar} \bigl[ H_{\rm tot} - \hbar \omega_{\ell} \hat{N}_{\rm tot}, \rho_{\rm tot}(t) \bigr],
\label{vonNeumann_eq}
\end{align}
where $\rho_{\rm tot}$ is the density operator of the total system in the rotating frame.
By applying the standard Born-Markov approximation \cite{BreuerPetruccione} to this equation,
we obtain a QME for the cQED:
\begin{align}
\frac{d}{dt} \rho_{\rm c}(t) = \mathcal{L} \rho_{\rm c}(t),
\label{QME_cQED}
\end{align}
where $\rho_{\rm c} = {\rm Tr}_{\rm b} {\rm Tr}_{\rm d} \rho_{\rm tot}$
is the density operator of the cQED,
and ${\rm Tr}_{\rm b}$ and ${\rm Tr}_{\rm d}$ are the traces over the Hilbert spaces of the baths and drain, respectively.
The Liouvillian $\mathcal{L}$ is composed of three parts:
\begin{align}
\mathcal{L}
&= \mathcal{L}_0 + \mathcal{L}_{\rm b} + \mathcal{L}_{\rm d},
\end{align}
where the superoperators $\mathcal{L}_0$ and $\mathcal{L}_{\rm d}$ act on any operator $\hat{O}_{\rm c}$ of the cQED as follows:
\begin{align}
\mathcal{L}_0 \hat{O}_{\rm c}
&= \frac{1}{i\hbar} \bigl[ H_{\rm c} + H_{\rm SB} - \hbar \omega_{\ell} N_{\rm c}, \hat{O}_{\rm c} \bigr],
\\
\mathcal{L}_{\rm d} \hat{O}_{\rm c}
&= - \frac{\kappa}{2} \bigl( \hat{c}^\dag \hat{c} \hat{O}_{\rm c} + \hat{O}_{\rm c} \hat{c}^\dag \hat{c} - 2 \hat{c} \hat{O}_{\rm c} \hat{c}^\dag \bigr),
\end{align}
and $\mathcal{L}_{\rm b}$ is the Liouvillian due to the baths (explicit form is irrelevant).
Here, $\kappa$ is the cavity loss rate.

In deriving the QME (\ref{QME_cQED}), we have made several assumptions.
The first one is weak couplings between the cQED and baths and between the cQED and drain,
so that we have used a second-order perturbation theory (Born approximation) in the couplings.
The second is that the time scale of the cQED is sufficiently slower than those of the baths and drain,
so that we have used the Markov approximation.
The third is that the drain is in the almost vacuum state.
The fourth one is the wide-band limit of the drain spectral function: $\Gamma_{\rm d}(\omega) = \kappa$ (constant).

In the present paper we consider the steady state of the cQED $\rho_{\rm c}^{\rm ss}$ in the rotating frame.
We assume that $\rho_{\rm c}^{\rm ss}$ is unique and stable;
i.e., the steady state is realized after sufficiently long time for any initial state.
Mathematically this means that zero is a non-degenerate eigenvalue of $\mathcal{L}$,
its corresponding eigenvector is $\rho_{\rm c}^{\rm ss}$,
and the real parts of the other eigenvalues are negative.
See Refs.~\cite{SpohnLebowitz, Frigerio, RivasHuelga} for conditions for $\mathcal{L}$ to have these properties.
If the QME (\ref{QME_cQED}) does not have these properties,
the cQED state after a long time might depend on the initial stateor it might oscillate among several states,
which is beyond the scope of the present paper.

\subsection{Radiation energy flow}

In the present setup, the cavity photons are leaked through the cavity loss due to the drain.
Thus we can define the radiation energy flow as the unit-time energy loss due to the drain:
\begin{align}
J_{\rm c \to d}^E \equiv - {\rm Tr_c} \bigl\{ (H_{\rm c} + H_{\rm SB}) \mathcal{L}_{\rm d} \rho_{\rm c}^{\rm ss} \bigr\},
\label{J_E_cd}
\end{align}
where ${\rm Tr_c}$ is the trace over the Hilbert space of the cQED.
In the present paper we use the sign convention
such that energy flows (including work and heat flows) are positive when delivered from the cQED to the drain.

As shown in Appendix~\ref{derivaton_energy_flow},
we can rewrite the above equation as
\begin{align}
J_{\rm c \to d}^E = \hbar\omega_\ell \kappa \langle \hat{c}^\dag \hat{c} \rangle_{\rm c}^{\rm ss}
+ \hbar \kappa \frac{d}{d\tau} {\rm Im} \bigl\langle \Check{c}^\dag(\tau) \hat{c} \bigr\rangle_{\rm c}^{\rm ss} \Big|_{\tau=0},
\label{J_E_cd_variant}
\end{align}
where $\Check{O}_{\rm c}(\tau) \equiv e^{\mathcal{L}^\dag \tau} \hat{O}_{\rm c}$ represents
the ``Heisenberg picture'' of any cQED operator $\hat{O}_{\rm c}$,
and $\langle \hat{O}_{\rm c} \rangle_{\rm c}^{\rm ss} \equiv {\rm Tr_c} \bigl( \rho_{\rm c}^{\rm ss} \hat{O}_{\rm c} \bigr)$
is the steady-state average.
Note that we define the adjoint $\mathcal{S}^\dag$ of a superoperator $\mathcal{S}$ such that
${\rm Tr}[\hat{O}_1^\dag \mathcal{S} \hat{O}_2] = {\rm Tr}[(\mathcal{S}^\dag \hat{O}_1)^\dag \hat{O}_2]$
holds for any pair of operators $\hat{O}_1, \hat{O}_2$.

\section{Quantum master equation for drain}

The purpose of the present paper is to decompose the radiation energy flow $J_{\rm c \to d}^E$ into the work and heat flows.
It is difficult, however,  to find a criterion for the decomposition from Eqs.~(\ref{J_E_cd}) and (\ref{J_E_cd_variant}).
Intuitively, the work and heat are respectively the ``systematic'' and ``random'' parts of energy transfer.
Therefore we expect that, for the decomposition,
it is necessary to investigate how the drain receives the energy flow.

To investigate this, we need to know the time evolution of the drain interacting with the cQED plus the baths.
Here we can use a QME for the drain to describe the time evolution
by regarding the cQED plus baths as an environment and by eliminating their degrees of freedom.
This is possible because we have assumed that the interaction between the cQED and drain is so weak that we can use a Born-type approximation.

The starting point in deriving a QME for the drain is the following QME for
the composite system composed of the cQED and drain:
\begin{align}
\frac{d}{dt} \rho_{\rm c+d}(t) &= (\mathcal{L} - \mathcal{L}_{\rm d} + \mathcal{K}_0 + \mathcal{M}_{\rm int}) \rho_{\rm c+d}(t),
\label{cQED+drain}
\end{align}
where $\rho_{\rm c+d}$ is the density operator of the composite system,
$\mathcal{K}_0 \hat{O}_{\rm c+d} = (1/i\hbar) \bigl[ H_{\rm d} - \hbar \omega_\ell \hat{N}_{\rm d}, \hat{O}_{\rm c+d} \bigr]$,
and $\mathcal{M}_{\rm int} \hat{O}_{\rm c+d} = (1/i\hbar) \bigl[ H_\text{c-d}, \hat{O}_{\rm c+d} \bigr]$
for any operator $\hat{O}_{\rm c+d}$ of the composite system.
We can derive the QME~(\ref{cQED+drain}) for the composite system as follows.
We first consider the cQED plus baths, and eliminate the baths' degrees of freedom in the standard Born-Markov approximation
to obtain a QME for the cQED.
After that, we connect it to the drain to obtain Eq.~(\ref{cQED+drain}).
We may justify the QME~(\ref{cQED+drain}) in the condition of weak coupling for $H_\text{c-d}$
(otherwise this is not justified \cite{NakataniOgawa}).
Indeed, if we eliminate the drain's degrees of freedom from Eq.~(\ref{cQED+drain}) in the Born-Markov approximation,
we recover the QME~(\ref{QME_cQED}) for the cQED.

To derive a QME for the drain from Eq.~(\ref{cQED+drain}), we eliminate the degrees of freedom of the cQED.
However, in contrast to the derivation of Eq.~(\ref{QME_cQED}), we cannot justify using the Markov approximation
in the derivation of a QME for the drain.
This is because we have assumed that the time scale of the drain is much faster than that of the cQED
in the derivation of Eq.~(\ref{QME_cQED}).
Therefore we have to describe the drain using a non-Markovian time-evolution equation.
One of the methods of describing the non-Markovian dynamics is a time-convolutionless (TCL) QME.
In particular, we employ a TCL-QME valid up to the second order in the coupling  between the cQED and drain.
The second order is necessary and sufficient to be consistent with
the coupling order in the QME~(\ref{QME_cQED}) for the cQED.
The second-order TCL-QME for the drain reads
\begin{align}
\frac{d}{dt} \rho_{\rm d}(t) &= \bigl[ \mathcal{K}_{\rm w} + \mathcal{K}_{\rm h}(t) \bigr] \rho_{\rm d}(t),
\label{TCLQME_drain}
\end{align}
where $\rho_{\rm d}(t) = {\rm Tr_c} \rho_{\rm c+d}(t)$ is the density operator of the drain at time $t$.
Here the superoperators are defined by the following equations:
\begin{widetext}
\begin{align}
\mathcal{K}_{\rm w} \hat{O}_{\rm d} &= \frac{1}{i\hbar} \bigl[ H_{\rm d} + \bar{H}_\text{c-d} - \hbar \omega_\ell \hat{N}_{\rm d}, \hat{O}_{\rm d} \bigr],
\label{Kw}
\\
\mathcal{K}_{\rm h}(t) \hat{O}_{\rm d} &= {\rm Tr_c} \int_{t_0}^t dt' \mathcal{M}'_{\rm int} e^{(\mathcal{L}+\mathcal{K}_0)(t-t')}
\mathcal{M}'_{\rm int} \bigl[ \rho_{\rm c}^{\rm ss} \otimes e^{-\mathcal{K}_0(t-t')} \hat{O}_{\rm d} \bigr],
\label{Kh}
\\
\mathcal{M}'_{\rm int} \hat{O}_{\rm c+d} &= \frac{1}{i\hbar} \bigl[ H_\text{c-d} - \bar{H}_\text{c-d}, \hat{O}_{\rm c+d} \bigr],
\end{align}
\end{widetext}
with any operators $\hat{O}_{\rm d}$ of the drain and $\hat{O}_{\rm c+d}$ of the composite system.
$\bar{H}_\text{c-d} \equiv {\rm Tr_c} ( \rho_{\rm c}^{\rm ss} H_\text{c-d} )
= \sum_k (\hbar g_k \langle \hat{c}^\dag \rangle_{\rm c}^{\rm ss} \hat{d}_k + \hbar g_k^* \langle \hat{c} \rangle_{\rm c}^{\rm ss} \hat{d}_k^\dag)$
is the steady-state-averaged interaction, and $t_0$ is the initial time.
In deriving the TCL-QME~(\ref{TCLQME_drain}), we have assumed that the initial state of the composite system is given by
$\rho_{\rm c+d}(t_0) = \rho_{\rm c}^{\rm ss} \otimes \rho_{\rm d}(t_0)$.
We shall show the detail of the derivation of the TCL-QME~(\ref{TCLQME_drain}) in Appendix~\ref{derivation_TCLQME}.

The TCL-QME~(\ref{TCLQME_drain}) is an effective equation of motion for only the drain
under the influence of the interaction with the cQED.
We note that that $\mathcal{K}_{\rm w}$ and $\mathcal{K}_{\rm h}$ respectively describe
the Hamiltonian and non-Hamiltonian dynamics of the drain.
As a result, we can split the influence of the interaction
into the systematic and random parts
%As a result, we can definitely split the systematic and random parts of the drain's time evolution
(due to $\mathcal{K}_{\rm w}$ and $\mathcal{K}_{\rm h}$, respectively).
Furthermore, Eq.~(\ref{Kw}) implies that the operator of the drain energy is $H_{\rm d} + \bar{H}_\text{c-d}$.

\section{Decomposition of radiation energy flow}

We now construct a method of decomposing the radiation energy flow into work and heat flows.
In subsection~\ref{subsec:definition1}, we define the work and heat flows
from the viewpoint of systematic and random interactions.
In the subsequent subsections, to justify the definition,
we analyze the energy flow in the rest frame (subsection~\ref{subsec:definition2})
and in terms of the entropy change (subsection~\ref{subsec:definition3}).
We show explicit forms of the work and heat flows, which are the main result of the present paper
in subsection~\ref{subsec:explicit_form},
and also show a relation to photoluminescence spectrum in subsection~\ref{subsec:PL_spectrum}.

\subsection{Definition of work and heat flows}
\label{subsec:definition1}

We have assumed that the drain is almost in the vacuum state in the derivation of the QME~(\ref{QME_cQED}) for the cQED.
Therefore it is reasonable to set the initial state of the drain to the vacuum state $| {\rm vac}\rangle$
in calculating the unit-time energy gain $J_{\rm d \leftarrow c}^E$ of the drain with the QME~(\ref{TCLQME_drain}) for the drain.
We thus define the energy flow $J_{\rm d \leftarrow c}^E$ by
\begin{align}
J_{\rm d \leftarrow c}^E \equiv {\rm Tr_d} \Bigl\{ (H_{\rm d} + \bar{H}_\text{c-d}) \bigl[ \mathcal{K}_{\rm w} + \mathcal{K}_{\rm h}(t) \bigr] \rho_{\rm d}(t) \Bigr\},
\end{align}
where $\rho_{\rm d}(t)$ is the solution of the QME~(\ref{TCLQME_drain})
with the initial condition $\rho_{\rm d}(t_0) = | {\rm vac}\rangle \langle {\rm vac}|$.
Note that we define $J_{\rm d \leftarrow c}^E$ independently of $J_{\rm c \to d}^E$ in Eq.~(\ref{J_E_cd})
(although we will show later that these are equal to each other).

We decompose this energy flow into two parts:
$J_{\rm d \leftarrow c}^E = J_{\rm d \leftarrow c}^{\rm work} + J_{\rm d \leftarrow c}^{\rm heat}$, where
\begin{align}
J_{\rm d \leftarrow c}^{\rm work} &\equiv {\rm Tr_d} \Bigl\{ (H_{\rm d} + \bar{H}_\text{c-d}) \mathcal{K}_{\rm w} \rho_{\rm d}(t) \Bigr\},
\label{work_flow}
\\
J_{\rm d \leftarrow c}^{\rm heat} &\equiv {\rm Tr_d} \Bigl\{ (H_{\rm d} + \bar{H}_\text{c-d}) \mathcal{K}_{\rm h}(t) \rho_{\rm d}(t) \Bigr\}.
\label{heat_flow}
\end{align}
We may regard $J_{\rm d \leftarrow c}^{\rm work}$ as the work flow because it is the energy gain due to the Hamiltonian (systematic) part of the equation of motion for the drain.
On the other hand, $J_{\rm d \leftarrow c}^{\rm heat}$ is the heat flow
because it is the energy gain due to the non-Hamiltonian (random) part.

Although this interpretation to identify the non-Hamiltonian part as heat
is consistent with that
in the thermodynamics of open quantum systems weakly coupled to thermal baths \cite{SpohnLebowitz},
it might seem rather intuitive.
We therefore justify this definition from different perspectives in the next two subsections.

\subsection{Justification 1: rest frame}
\label{subsec:definition2}

We may justify this decomposition also from the viewpoint of the rest frame as follows.
In the rest frame the Hamiltonian (energy operator) of the drain
$\hat{H}_{\rm d}^{\rm rest}(t) = e^{-i\omega_\ell t\hat{N}_{\rm d}} (H_{\rm d} + \bar{H}_\text{c-d}) e^{i\omega_\ell t\hat{N}_{\rm d}}$
is modulated in time through the photonic amplitude $\langle \hat{c} \rangle_{\rm c}^{\rm ss} e^{-i\omega_\ell t}$ of the cQED.
Hence we may interpret the cQED as an external agent which varies the control parameter in the drain.
Then the unit-time change of the drain energy (which is equal to $J_{\rm d \leftarrow c}^E$) has two contributions:
\begin{align}
J_{\rm d \leftarrow c}^E &= {\rm Tr_d} \left\{ \hat{\rho}_{\rm d}^{\rm rest}(t) \frac{d}{dt} \hat{H}_{\rm d}^{\rm rest}(t) \right\}
\notag\\
&+ {\rm Tr_d} \left\{ \hat{H}_{\rm d}^{\rm rest}(t) \frac{d}{dt} \hat{\rho}_{\rm d}^{\rm rest}(t) \right\},
\end{align}
where $\hat{\rho}_{\rm d}^{\rm rest}(t) = e^{-i\omega_\ell t\hat{N}_{\rm d}} \rho_{\rm d}(t) e^{i\omega_\ell t\hat{N}_{\rm d}}$
is the drain's state in the rest frame.
The first term is attributed to the controllable change of the Hamiltonian itself,
whereas the second is attributed to the uncontrollable change of the drain's state.
Therefore we may interpret the first and second terms as the changes in work and heat, respectively.
In fact, as shown in Appendix~\ref{proof_equivalence}, the following relations hold:
\begin{align}
J_{\rm d \leftarrow c}^{\rm work} &= {\rm Tr_d} \left\{ \hat{\rho}_{\rm d}^{\rm rest}(t) \frac{d}{dt}\hat{H}_{\rm d}^{\rm rest}(t) \right\},
\label{equivalence_work}
\\
J_{\rm d \leftarrow c}^{\rm heat} &= {\rm Tr_d} \left\{ \hat{H}_{\rm d}^{\rm rest}(t) \frac{d}{dt}\hat{\rho}_{\rm d}^{\rm rest}(t) \right\}.
\label{equivalence_heat}
\end{align}
We also note that this interpretation is consistent
with those in the statistical mechanics \cite{Davidson},
the open quantum systems \cite{SpohnLebowitz, Alicki1979, Kosloff1984},
and the thermodynamics of small systems \cite{Sekimoto, BustamanteLiphardtRitort}.

\subsection{Justification 2: entropy}
\label{subsec:definition3}

We further justify the decomposition in terms of entropy.
For this purpose we investigate the time derivative of the von Neumann entropy
$S_{\rm vN}( \rho_{\rm d} ) \equiv -{\rm Tr_d} \rho_{\rm d} \ln \rho_{\rm d}$ of the drain ($k_{\rm B}=1$):
\begin{align}
\frac{d}{dt} S_{\rm vN}\bigl( \rho_{\rm d}(t) \bigr)
&= -{\rm Tr_d} \bigl\{ \bigl( d\rho_{\rm d}(t)/dt \bigr) \ln \rho_{\rm d}(t) \bigr\}
\notag\\
&= -{\rm Tr_d} \Bigl\{ \bigl( \ln \rho_{\rm d}(t) \bigr)
\bigl[ \mathcal{K}_{\rm w} + \mathcal{K}_{\rm h}(t) \bigr] \rho_{\rm d}(t) \Bigr\}
\notag\\
&= -{\rm Tr_d} \Bigl\{ \bigl(\ln \rho_{\rm d}(t) \bigr) \mathcal{K}_{\rm h}(t) \rho_{\rm d}(t) \Bigr\}.
\label{entropy_flow}
\end{align}
We have used the trace-preserving property of the QME in the first line,
the form of the QME~(\ref{TCLQME_drain}) in the second line.
In the third line,
we have used the fact that the Hamiltonian dynamics does not change the von Neumann entropy,
$-{\rm Tr_d} \bigl\{ \bigl(\ln \rho_{\rm d}(t) \bigr) \mathcal{K}_{\rm w} \rho_{\rm d}(t) \bigr\} = 0$.
This fact used in the third line means that the work flow $J_{\rm d \leftarrow c}^{\rm work}$
does not induce the entropy change in the drain.

We here note that it is reasonable to evaluate Eq.~(\ref{entropy_flow})
on the second order in the coupling between the cQED and drain
because the QME~(\ref{TCLQME_drain}) is valid up to the second order.
Furthermore, since $\mathcal{K}_{\rm h}$ in Eq.~(\ref{entropy_flow}) is on the second order,
we can evaluate $\rho_{\rm d}(t)$ on the zeroth order as
$\rho_{\rm d}(t) \simeq e^{(H_{\rm d} - \hbar \omega_\ell \hat{N}_{\rm d}) (t-t_0)/i\hbar}
| {\rm vac} \rangle \langle {\rm vac} | e^{-(H_{\rm d} - \hbar \omega_\ell \hat{N}_{\rm d}) (t-t_0)/i\hbar}
= | {\rm vac} \rangle \langle {\rm vac} |$.
However, if we apply this evaluation of $\rho_{\rm d}$,
Eq.~(\ref{entropy_flow}) will diverge due to the term of $\ln \rho_{\rm d}$.
This is because the vacuum state is the zero-temperature equilibrium state of the drain.
Then, to investigate its diverging behavior,
we employ $\rho_{\rm d} \simeq \lim_{T_{\rm d} \to 0} e^{-H_{\rm d}/T_{\rm d}}/Z_{\rm d}$
when evaluating $\ln \rho_{\rm d}$ (where $Z_{\rm d} = {\rm Tr_d} e^{-H_{\rm d}/T_{\rm d}}$).
By using the above evaluation, we rewrite Eq.~(\ref{entropy_flow}) as
\begin{align}
\frac{d}{dt} S_{\rm vN}\bigl( \rho_{\rm d}(t) \bigr)
\simeq \lim_{T_{\rm d} \to 0} \frac{1}{T_{\rm d}} {\rm Tr_d} \bigl\{ H_{\rm d} \mathcal{K}_{\rm h}(t) | {\rm vac} \rangle \langle {\rm vac} | \bigr\},
\label{entropy_flow_2nd}
\end{align}
where we have again used the trace-preserving property of $\mathcal{K}_{\rm h}$.

We can also evaluate the heat flow $J_{\rm d \leftarrow c}^{\rm heat}$ on the second order in the coupling as
$J_{\rm d \leftarrow c}^{\rm heat} \simeq
{\rm Tr_d} \bigl\{ H_{\rm d} \mathcal{K}_{\rm h}(t) | {\rm vac} \rangle \langle {\rm vac} | \bigr\}$
[see Eq.~(\ref{J_heat_2nd}) in Appendix~\ref{derivation_work_heat_flows}].
By comparing this equation with Eq.~(\ref{entropy_flow_2nd}),
we conclude that the heat flow $J_{\rm d \leftarrow c}^{\rm heat}$
is proportional to the unit-time entropy change in the drain:
\begin{align}
\lim_{T_{\rm d} \to 0} \frac{J_{\rm d \leftarrow c}^{\rm heat}}{T_{\rm d}}
= \frac{d}{dt} S_{\rm vN}\bigl( \rho_{\rm d}(t) \bigr).
\label{heat_entropy}
\end{align}
This result justifies the definition of $J_{\rm d \leftarrow c}^{\rm heat}$ and $J_{\rm d \leftarrow c}^{\rm work}$.
That is, our definition is consistent with the following statement of the thermodynamics:
the heat is the energy transfer accompanied by the entropy transfer, and the work is not accompanied by it.

Conversely, the relation~(\ref{heat_entropy}) implies the following two notions.
One is that the von Neumann entropy $S_{\rm vN}\bigl( \rho_{\rm d}(t) \bigr)$ is regarded as
the thermodynamic entropy of the drain.
This is consistent with the assumption that the drain is almost in its (zero-temperature) equilibrium state.
The other is that the energy (work and heat) transfer is quasi-static for the drain
because the equality (not inequality) holds in Eq.~(\ref{heat_entropy}).
This is consistent with the assumption that the drain is ``almost always'' the same state (vacuum state).

\subsection{Explicit forms of work and heat flows}
\label{subsec:explicit_form}

As shown in Appendix~\ref{derivation_work_heat_flows}, we can write the flows in more explicit forms:
\begin{align}
J_{\rm d \leftarrow c}^{\rm work} &= \hbar \omega_\ell \kappa \big| \langle \hat{c} \rangle_{\rm c}^{\rm ss} \big|^2,
\label{work_flow_explicit}
\\
J_{\rm d \leftarrow c}^{\rm heat} &= \hbar \omega_\ell \kappa \langle \delta\hat{c}^\dag \delta\hat{c} \rangle_{\rm c}^{\rm ss}
+ \hbar \kappa \frac{d}{d\tau} {\rm Im} \langle \Check{c}^\dag(\tau) \hat{c} \rangle_{\rm c}^{\rm ss} \Big|_{\tau=0},
\label{heat_flow_explicit}
\end{align}
where $\delta \hat{c} \equiv \hat{c} - \langle \hat{c} \rangle_{\rm c}^{\rm ss}$.
These forms, which give a practical method of the decomposition, are the main result of the present paper.
Here we make four remarks on this result.

The first one is that the energy conservation law does hold.
We can easily show this law in the form of energy balance equation, $J_{\rm c \to d}^E = J_{\rm d \leftarrow c}^E$,
by comparing $J_{\rm c \to d}^E$ given by Eq.~(\ref{J_E_cd_variant})
and $J_{\rm d \leftarrow c}^E = J_{\rm d \leftarrow c}^{\rm work} + J_{\rm d \leftarrow c}^{\rm heat}$
with Eqs.~(\ref{work_flow_explicit}) and (\ref{heat_flow_explicit}).
This result supports the utilization of the QME~(\ref{TCLQME_drain}) in defining the work and heat flows.
We also note that this law is a local version of the first law of thermodynamics,
in the sense that this law is associated with the local interface between the cQED and drain.
(Of course, the global energy conservation law also holds in the present setup.)

The second is that the work and heat correspond respectively to coherent (finite photon amplitude) and incoherent parts of the energy transfer
from the cQED to the drain.
This gives an intuitive understanding of the decomposition.

The third is that the explicit forms [Eqs.~(\ref{work_flow_explicit}) and (\ref{heat_flow_explicit})] are independent of time $t$
although they seem to be dependent on $t$ in the definitions [Eqs.~(\ref{work_flow}) and (\ref{heat_flow})].
This result is consistent with the fact that we are investigating a continuous heat engine in the steady-state situation of the cQED.

The final remark is that we need not use the QME~(\ref{TCLQME_drain}) in the practical calculation of work and heat flows.
From the QME~(\ref{QME_cQED}) for the cQED,
we can obtain all the information necessary to calculate them by using Eqs.~(\ref{work_flow_explicit}) and (\ref{heat_flow_explicit}).

\subsection{Relation to photoluminescence spectrum}
\label{subsec:PL_spectrum}

We furthermore find relations between the (total, work, and heat) energy flows and the photoluminescence spectrum.

We define the photoluminescence spectrum $I(\omega)$ of the steady-state cQED as
\begin{align}
I(\omega) \equiv \frac{\kappa}{\pi} {\rm Re} \int_0^\infty d\tau
\langle \Check{c}^\dag(\tau) \hat{c} \rangle_{\rm c}^{\rm ss} e^{-i(\omega - \omega_\ell) \tau}.
\end{align}
Here, the factor $e^{i \omega_\ell \tau}$ appears because the steady state and $\Check{c}^\dag(\tau)$ are in the rotating frame.
The spectrum $I(\omega)$ corresponds to the steady-state average of the unit-time and unit-frequency photon number observed in the drain.
The Fourier transform of the above equation gives the correlation function:
\begin{align}
\langle \Check{c}^\dag(\tau) \hat{c} \rangle_{\rm c}^{\rm ss}
= \frac{1}{\kappa} \int_{-\infty}^\infty d\omega I(\omega) e^{i(\omega - \omega_\ell) \tau}.
\end{align}
By using this equation, we rewrite the individual terms on the right-hand side of Eq.~(\ref{J_E_cd_variant}) as
\begin{align}
\hbar\omega_\ell \kappa \langle \hat{c}^\dag \hat{c} \rangle_{\rm c}^{\rm ss}
&= \hbar\omega_\ell \int_{-\infty}^\infty d\omega I(\omega),
\\
\hbar\kappa \frac{d}{d\tau} \langle \Check{c}^\dag(\tau) \hat{c} \rangle_{\rm c}^{\rm ss} \big|_{\tau = 0}
&= \int_{-\infty}^\infty d\omega (\hbar\omega - \hbar\omega_\ell) I(\omega).
\end{align}
We thus obtain the relation between the total energy flow $J_{\rm c \to d}^E$ and the spectrum $I(\omega)$:
\begin{align}
J_{\rm c \to d}^E = \int_{-\infty}^\infty d\omega \hbar \omega I(\omega).
\end{align}
We may intuitively understand this relation:
the sum of the energy $\hbar \omega$ multiplied by its radiation strength $I(\omega)$ gives the energy flow $J_{\rm c \to d}^E$.

Moreover, we may decompose the spectrum into the coherent peak and the rest (incoherent part) as
$I(\omega) = I_{\rm coh}(\omega) + I_{\rm inc}(\omega)$, where
\begin{align}
I_{\rm coh}(\omega) &= \kappa \big| \langle \hat{c} \rangle_{\rm c}^{\rm ss} \big|^2 \delta(\omega - \omega_\ell),
\\
I_{\rm inc}(\omega) &= \frac{\kappa}{\pi} {\rm Re} \int_0^\infty d\tau
\langle \delta\Check{c}^\dag(\tau) \delta\hat{c} \rangle_{\rm c}^{\rm ss} e^{-i(\omega - \omega_\ell) \tau}.
\end{align}
From these equations with Eqs.~(\ref{work_flow_explicit}) and (\ref{heat_flow_explicit}),
we obtain the relations between work and heat flows and the spectra:
\begin{align}
J_{\rm d \leftarrow c}^{\rm work} &= \int_{-\infty}^\infty d\omega \hbar \omega I_{\rm coh}(\omega),
\label{work_flow_spectrum}
\\
J_{\rm d \leftarrow c}^{\rm heat} &= \int_{-\infty}^\infty d\omega \hbar \omega I_{\rm inc}(\omega).
\label{heat_flow_spectrum}
\end{align}
Again, we may intuitively understand these relations:
the sum of the energy $\hbar \omega$ multiplied by its coherent (incoherent) radiation strength $I_{\rm coh}(\omega)$ [$I_{\rm inc}(\omega)$]
gives the work (heat) flow $J_{\rm d \leftarrow c}^{\rm work}$ ($J_{\rm d \leftarrow c}^{\rm heat}$).

\section{Concluding remarks}\label{sec:Conclusion}

In the present paper we have proposed a method, Eqs.~(\ref{work_flow_explicit}) and (\ref{heat_flow_explicit}),
of decomposing the radiation energy into the work and heat.
This method provides with a reliable foundation for the intuition that the coherent part is the work while the incoherent part is the heat.
The key point in deriving this method is to explicitly consider the time evolution of the drain in the QME~ (\ref{TCLQME_drain})
to investigate in what forms the drain receives the radiation energy from the cQED.

We note that the relevant assumptions in the setup for our formulation are:
(i) the cavity is single mode,
(ii) the drain is a system of the free electromagnetic field ($\hat{H}_{\rm d} = \sum_k \hbar \omega_k \hat{d}_k^\dag \hat{d}_k$)
and is in almost vacuum state,
(iii) the interaction between the cQED and drain is weak and given by
$\hat{H}_\text{c-d} = \sum_k (\hbar g_k \hat{c}^\dag \hat{d}_k + \hbar g_k^* \hat{c} \hat{d}_k^\dag)$,
(iv) the baths interact only with the matter part of the cQED,
(v) the photonic amplitude rotates with a single frequency (like a single-mode laser),
and (vi) the excitation number conserves without the SB field.
In this sense, the decomposition method in the present paper is rather general and widely applicable
to cavity QED and circuit QED systems.
For example, the method will shed a new light on the laser as a quantum heat engine;
we should revisit the thermodynamic efficiency and power of the laser
by counting not all the output energy but the pure work.

Extending the method beyond the above assumptions is a future issue.
Extension to multimode cases [concerning (i) and (v)] is of particular importance.
Also, extension to cases of strong interaction between the cQED and drain would be of interest.
We expect that one could formulate a method similar to the present one
by using the singular coupling limit \cite{GoriniKossakowski, FrigerioGorini}
or the hierarchal equations of motion approach \cite{Tanimura, KatoTanimura}.

Another future issue concerns the SB field.
In theoretical situations without the SB field,
we cannot calculate $J_{\rm d \leftarrow c}^{\rm work}$ with Eq.~(\ref{work_flow_explicit})
since $\langle \hat{c} \rangle_{\rm c}^{\rm ss}$ vanishes.
One promising way to extend the decomposition method to such cases
is to use the relations to the photoluminescence spectrum $I(\omega)$.
The spectrum usually has a very sharp peak above the laser threshold even without the SB field
although it is not a strict delta function but has a narrow width due to dephasing.
This fact would allow us to decompose the spectrum into coherent and incoherent parts, $I_{\rm coh}(\omega)$ and $I_{\rm inc}(\omega)$.
Then we may calculate the work and heat flows with Eqs.~(\ref{work_flow_spectrum}) and (\ref{heat_flow_spectrum}).
This also suggests that one may estimate work and heat flows in experiments by measuring $I(\omega)$.
At present, however, it is not clear how to unambiguously define the coherent and incoherent parts of the spectrum.

Related to the above, in Ref.~\cite{Gelbwaser-Klimovsky_etal},
they introduced additional degrees of freedom (called ``Piston'') instead of the SB field.
The Piston is placed at the interface of the system and the external oscillator
(the cQED and the drain in our case).
They showed that the initial state of the Piston must be nonpassive to obtain the work.
Further investigation without the SB field in our setup would clarify
whether the Piston (and its nonpassive initial state) is essential for the work extraction.

As another related study, we refer to Ref.~\cite{Weimer_etal},
where they provided a similar decomposition of work and heat.
There, to define work and heat, they employed a local effective equation of motion for a subsystem
interacting with another subsystem.
They introduced an effective local Hamiltonian by averaging the interaction with the other subsystem
(like $\bar{H}_\text{c-d}$ in the present paper)
and incorporated a part of it in defining the work.
Investigation of consistency between the decompositions in Ref.~\cite{Weimer_etal} and the present paper
is also a future issue.
Since the interaction is not necessarily weak in the method in Ref.~\cite{Weimer_etal},
it may be also useful to extend our result to the strong coupling regime.

One might suppose that the ultimate origin of the work to the drain is only the work done by the SB field
because the SB field seems to be the only work source of the total system.
However this is not the case.
This is because the cQED plays the role of a thermal machine.
Therefore a part of the heat from the (hot) heat baths is converted to a part of the work to the drain
if it operates as a heat engine,
and a part of the work from the SB field is converted to a part of heat to the baths
if it operates as a heat pump or a refrigerator.
The QME (\ref{QME_cQED}) for the cQED guarantees only the total energy conservation:
$J_{\rm c \to d}^E = J_{\rm SB \to c}^{\rm work} + J_{\rm b \to c}^{\rm heat}$,
where $J_{\rm c \to d}^E$ is the energy flow from the cQED to the drain [Eq.~(\ref{J_E_cd})],
$J_{\rm SB \to c}^{\rm work} \equiv {\rm Tr_c} \bigl\{ (H_{\rm c} + H_{\rm SB}) \mathcal{L}_0 \rho_{\rm c}^{\rm ss} \bigr\}$
is the work flow from the SB field to the cQED, and
$J_{\rm b \to c}^{\rm heat} \equiv {\rm Tr_c} \bigl\{ (H_{\rm c} + H_{\rm SB}) \mathcal{L}_{\rm b} \rho_{\rm c}^{\rm ss} \bigr\}$
is the sum of the heat flows from the heat baths.
Combining this energy balance equation with the decomposition
$J_{\rm c \to d}^E = J_{\rm d \leftarrow c}^{\rm work} + J_{\rm d \leftarrow c}^{\rm heat}$,
we obtain $J_{\rm d \leftarrow c}^{\rm work} + J_{\rm d \leftarrow c}^{\rm heat}
= J_{\rm SB \to c}^{\rm work} + J_{\rm b \to c}^{\rm heat}$.
However this does not mean the equivalence of the individual terms.
Indeed, we can clearly show $J_{\rm d \leftarrow c}^{\rm work} \neq J_{\rm SB \to c}^{\rm work}$
by comparing Eq.~(\ref{work_flow_explicit})
with an explicit form of the work flow from the SB field,
$J_{\rm SB \to c}^{\rm work} = - \hbar \omega_\ell {\rm Im} \bigl\{2 f_{\rm SB}^* \langle \hat{c} \rangle_{\rm c}^{\rm ss} / \hbar \bigr\}$.

We also mention the difference between the results in Ref.~\cite{HorowitzEsposito} and the present paper.
In Ref.~\cite{HorowitzEsposito},
a system interacts with a reservoir driven by an external force
and exchanges both work and heat with the reservoir.
The situation of this system seems similar to that of the drain in the present setup.
However there exists a crucial difference:
In Ref.~\cite{HorowitzEsposito} the state of the reservoir is assumed to be described by the generalized Gibbs ensemble,
whereas the state of the environment for the drain (the driven cQED plus heat baths) is given by $\rho_{\rm c}^{\rm ss}$.
As a result, the conclusions are also different from each other.
In Ref.~\cite{HorowitzEsposito} the work from the reservoir to the system originates only from that by the external force,
whereas the work from the cQED to the drain originates both the work from SB field and the heat from the baths as mentioned in the previous paragraph.

Finally, we make a comment on a relationship between our definition of heat
and the non-work in the sense of Ref.~\cite{HatsopoulosGyftopoulos}.
Only from the arguments in subsections \ref{subsec:definition1} and \ref{subsec:definition2},
$J_{\rm d \leftarrow c}^{\rm work}$ may be regarded as non-work (but not necessarily as heat)
in the sense of Ref.~\cite{HatsopoulosGyftopoulos}.
However, because the argument in the subsection \ref{subsec:definition3}
connects $J_{\rm d \leftarrow c}^{\rm work}$ to the entropy flow,
$J_{\rm d \leftarrow c}^{\rm work}$ may be regarded as heat flow
even in the sense of Ref.~\cite{HatsopoulosGyftopoulos}.
Related to this comment, we refer to a statement in Ref.~\cite{BerettaGyftopoulos}
that the interaction through radiation between two black bodies at different temperatures
is not heat but non-work.
In our definition, on the other hand, the black body radiation is heat.
We think that this discrepancy between Ref.~\cite{BerettaGyftopoulos} and ours
originates from the difference in the situations.
In the present paper,
we concentrate our interest on the emission of photons (and accompanying energy and entropy flows)
at the local interface between the cQED and the drain (free space of electromagnetic field).
In Ref.~\cite{BerettaGyftopoulos}, by contrast, they are interested in not only the emission
but also the absorption of the emitted photons after they travel across the drain.

\begin{acknowledgments}
The authors acknowledge M. Bamba, K. Kamide, and Y. Yamada for their helpful advice.
This work was supported by JSPS KAKENHI (Grants No. 26287087)
and by ImPACT Program of Council for Science, Technology and Innovation (Cabinet Office, Government of Japan).
\end{acknowledgments}

\appendix

\section{Derivation of Eq.~(\ref{J_E_cd_variant})}
\label{derivaton_energy_flow}

From the definition of $J_{\rm c \to d}^E$ [Eq.~(\ref{J_E_cd})], we have
\begin{align}
&J_{\rm c \to d}^E = - {\rm Tr_c} \bigl\{ \rho_{\rm c}^{\rm ss} \mathcal{L}_{\rm d}^\dag (H_{\rm c} + H_{\rm SB}) \bigr\}
\notag\\
&= \frac{\kappa}{2} {\rm Tr_c} \Bigl\{ \rho_{\rm c}^{\rm ss}
\Bigl( \hat{c}^\dag \bigr[ \hat{c}, H_{\rm c} + H_{\rm SB} \bigr] - \bigl[ \hat{c}^\dag, H_{\rm c} + H_{\rm SB} \bigr] \hat{c} \Bigr) \Bigr\}.
\label{J_E_cd_App}
\end{align}
We rewrite $[ \hat{c}, H_{\rm c} + H_{\rm SB} ] $ in the above equation as
\begin{align}
\bigr[ \hat{c}, H_{\rm c} + H_{\rm SB} \bigr]
&= i \hbar \mathcal{L}_0^\dag \hat{c} + [\hat{c}, \hbar \omega_\ell N_{\rm c}]
\notag\\
&= i \hbar (\mathcal{L}^\dag - \mathcal{L}_{\rm d}^\dag) \hat{c} + \hbar \omega_\ell \hat{c}
\notag\\
&= i \hbar \mathcal{L}^\dag \hat{c} + \frac{i \hbar \kappa}{2} \hat{c} + \hbar \omega_\ell \hat{c},
\end{align}
where we have used $\mathcal{L}_{\rm b}^\dag \hat{c} = 0$ (because the baths interact only with the matter part of the cQED) in the second line
and $\mathcal{L}_{\rm d}^\dag \hat{c} = -\kappa \hat{c} /2$ in the third line.
Similarly, we rewrite $[ \hat{c}^\dag, H_{\rm c} + H_{\rm SB} ] $ as
\begin{align}
\bigr[ \hat{c}^\dag, H_{\rm c} + H_{\rm SB} \bigr]
= i \hbar \mathcal{L}^\dag \hat{c}^\dag + \frac{i \hbar \kappa}{2} \hat{c}^\dag  - \hbar \omega_\ell \hat{c}^\dag.
\end{align}
By substituting these equations into Eq.~(\ref{J_E_cd_App}), we obtain
\begin{align}
&J_{\rm c \to d}^E = \hbar\omega_\ell \kappa \langle \hat{c}^\dag \hat{c} \rangle_{\rm c}^{\rm ss}
+ \frac{i \hbar \kappa}{2} {\rm Tr_c} \Bigl\{ \rho_{\rm c}^{\rm ss}
\Bigl( \hat{c}^\dag \mathcal{L}^\dag \hat{c} - (\mathcal{L}^\dag \hat{c}^\dag) \hat{c} \Bigr) \Bigr\}
\notag\\
&= \hbar\omega_\ell \kappa \langle \hat{c}^\dag \hat{c} \rangle_{\rm c}^{\rm ss}
+ \frac{i \hbar \kappa}{2} \frac{d}{d\tau} {\rm Tr_c} \Bigl\{ \rho_{\rm c}^{\rm ss}
\Bigl( \hat{c}^\dag \Check{c}(\tau) - \Check{c}^\dag(\tau) \hat{c} \Bigr) \Bigr\} \Big|_{\tau=0}
\notag\\
&= \hbar\omega_\ell \kappa \langle \hat{c}^\dag \hat{c} \rangle_{\rm c}^{\rm ss}
+ \frac{i \hbar \kappa}{2} \frac{d}{d\tau}
\Bigl( \langle \Check{c}^\dag(\tau) \hat{c} \rangle_{\rm c}^{\rm ss *}
- \langle \Check{c}^\dag(\tau) \hat{c} \rangle_{\rm c}^{\rm ss} \Bigr) \Big|_{\tau=0}.
\end{align}
We thus derive Eq.~(\ref{J_E_cd_variant}).

\section{Derivation of Eqs.~(\ref{equivalence_work}) and (\ref{equivalence_heat})}
\label{proof_equivalence}

We can derive Eq.~(\ref{equivalence_work}) as follows:
\begin{align}
&{\rm Tr_d} \left\{ \hat{\rho}_{\rm d}^{\rm rest}(t) \frac{d}{dt} \hat{H}_{\rm d}^{\rm rest}(t) \right\}
\notag\\
&= {\rm Tr_d} \left\{ \hat{\rho}_{\rm d}^{\rm rest}(t) \frac{1}{i\hbar} \bigl[ \hbar \omega_\ell \hat{N}_{\rm d}, \hat{H}_{\rm d}^{\rm rest}(t) \bigr] \right\}
\notag\\
&= {\rm Tr_d} \left\{ \rho_{\rm d}(t) \frac{1}{i\hbar} \bigl[ \hbar \omega_\ell \hat{N}_{\rm d}, H_{\rm d} + \bar{H}_\text{c-d} \bigr] \right\}
\notag\\
&= -{\rm Tr_d} \left\{ \rho_{\rm d}(t) \frac{1}{i\hbar} \bigl[ H_{\rm d} + \bar{H}_\text{c-d} - \hbar \omega_\ell \hat{N}_{\rm d}, H_{\rm d} + \bar{H}_\text{c-d} \bigr] \right\}
\notag\\
&= {\rm Tr_d} \bigl\{ \rho_{\rm d}(t) \mathcal{K}_{\rm w}^\dag ( H_{\rm d} + \bar{H}_\text{c-d} ) \bigr\}
\notag\\
&= J_{\rm d \leftarrow c}^{\rm work}.
\label{J_work_rest}
\end{align}

We can derive Eq.~(\ref{equivalence_heat}) as follows:
\begin{align}
{\rm Tr_d} & \left\{ \hat{H}_{\rm d}^{\rm rest}(t) \frac{d}{dt} \hat{\rho}_{\rm d}^{\rm rest}(t) \right\}
\notag\\
=&~ {\rm Tr_d} \left\{ \hat{H}_{\rm d}^{\rm rest}(t) \frac{d}{dt} \left( e^{-i \omega_\ell t \hat{N}_{\rm d}} \rho_{\rm d}(t)
e^{i \omega_\ell t \hat{N}_{\rm d}} \right) \right\}
\notag\\
=&~ {\rm Tr_d} \left\{ \hat{H}_{\rm d}^{\rm rest}(t) e^{-i \omega_\ell t \hat{N}_{\rm d}} \left( \frac{d}{dt} \rho_{\rm d}(t) \right)
e^{i \omega_\ell t \hat{N}_{\rm d}} \right\}
\notag\\
&+ {\rm Tr_d} \left\{ \hat{H}_{\rm d}^{\rm rest}(t) \frac{1}{i\hbar} \bigl[ \hbar \omega_\ell \hat{N}_{\rm d}, \hat{\rho}_{\rm d}^{\rm rest}(t) \bigr] \right\}
\notag\\
=&~ {\rm Tr_d} \bigl\{ (H_{\rm d} + \bar{H}_\text{c-d}) \bigl[ \mathcal{K}_{\rm w} + \mathcal{K}_{\rm h}(t) \bigr] \rho_{\rm d}(t) \bigr\}
\notag\\
&- {\rm Tr_d} \bigl\{ (H_{\rm d} + \bar{H}_\text{c-d}) \mathcal{K}_{\rm w} \rho_{\rm d}(t) \bigr\}
\notag\\
=&~ {\rm Tr_d} \bigl\{ (H_{\rm d} + \bar{H}_\text{c-d}) \mathcal{K}_{\rm h}(t) \rho_{\rm d}(t) \bigr\}
\notag\\
=&~ J_{\rm d \leftarrow c}^{\rm heat}.
\end{align}
We have used that the term in the fourth line is equal to
${\rm Tr_d} \bigl\{ \hat{H}_{\rm d}^{\rm rest}(t) (1/i\hbar) [ \hbar \omega_\ell \hat{N}_{\rm d}, \hat{\rho}_{\rm d}^{\rm rest}(t) ] \bigr\}
=- {\rm Tr_d} \bigl\{ (1/i\hbar) [ \hbar \omega_\ell \hat{N}_{\rm d}, \hat{H}_{\rm d}^{\rm rest}(t) ] \hat{\rho}_{\rm d}^{\rm rest}(t) \bigr\}
= -J_{\rm d \leftarrow c}^{\rm work}$.

\section{Derivation of explicit forms of work and heat flows}
\label{derivation_work_heat_flows}

We here derive the explicit forms of work and heat flows, Eqs.~(\ref{work_flow_explicit}) and (\ref{heat_flow_explicit}).
At several points in the derivation we use the order estimation with respect to the coupling strength between the cQED and drain.
Note that we should estimate on the second order in the coupling because the QME~(\ref{TCLQME_drain}) is valid up to the second order.
We also use the wide-band limit of the drain spectral function, $\Gamma_{\rm d}(\omega) = \kappa$ (constant),
which is consistent with the fact that we have derived the QME~(\ref{QME_cQED}) in this limit.

\subsection{Derivation of Eq.~(\ref{work_flow_explicit})}

First we derive the explicit form of the work flow $J_{\rm d \leftarrow c}^{\rm work}$.
From the definition of $J_{\rm d \leftarrow c}^{\rm work}$ [Eq.~(\ref{work_flow})], we have
\begin{align}
J_{\rm d \leftarrow c}^{\rm work}
= {\rm Tr_d} \bigl\{ \rho_{\rm d}(t) \mathcal{K}_{\rm w}^\dag ( H_{\rm d} + \bar{H}_\text{c-d}) \bigr\}.
\label{work_flow_2}
\end{align}
Here, $\mathcal{K}_{\rm w}^\dag \bigl( H_{\rm d} + \bar{H}_\text{c-d} \bigr)$
in the above equation becomes
\begin{align}
&\mathcal{K}_{\rm w}^\dag \Bigl( H_{\rm d} + \bar{H}_\text{c-d} \bigr)
\notag\\
&= -\frac{1}{i\hbar} \bigl[ H_{\rm d} + \bar{H}_\text{c-d} - \hbar \omega_\ell \hat{N}_{\rm d} ,
H_{\rm d} + \bar{H}_\text{c-d} \bigr]
\notag\\
&= \frac{1}{i\hbar} \bigl[ \hbar \omega_\ell \hat{N}_{\rm d} , \bar{H}_\text{c-d} \bigr]
\notag\\
&= i \hbar \omega_\ell \sum_k \bigl( g_k \langle \hat{c}^\dag \rangle_{\rm c}^{\rm ss} \hat{d}_k
- g_k^* \langle \hat{c} \rangle_{\rm c}^{\rm ss} \hat{d}_k^\dag \bigr).
\label{KwHd}
\end{align}
This equation is on the first order in the coupling.
Therefore it is sufficient to evaluate $\rho_{\rm d}(t)$ in Eq.~(\ref{work_flow_2}) on the first order in the coupling:
$\rho_{\rm d}(t) \simeq e^{\mathcal{K}_{\rm w} (t-t_0)} | {\rm vac} \rangle \langle {\rm vac} |
= e^{H_{\rm d}^{\rm eff} (t-t_0)/i\hbar} | {\rm vac} \rangle \langle {\rm vac} | e^{-H_{\rm d}^{\rm eff} (t-t_0)/i\hbar}$,
where $H_{\rm d}^{\rm eff} \equiv H_{\rm d} + \bar{H}_\text{c-d} - \hbar \omega_\ell \hat{N}_{\rm d}$.
We can rigorously calculate $e^{H_{\rm d}^{\rm eff} (t-t_0)/i\hbar} | {\rm vac} \rangle$ as
\begin{align}
e^{H_{\rm d}^{\rm eff} (t-t_0)/i\hbar} | {\rm vac} \rangle = \bigotimes_k C_k(t) | \gamma_k(t) \rangle.
\label{coherent_state}
\end{align}
Here $C_k(t)$ is an unimportant phase factor and
$| \gamma_k(t) \rangle$ is the coherent state having the complex amplitude
$\gamma_k(t) = g_k^* \langle \hat{c} \rangle_{\rm c}^{\rm ss} (e^{-i \Omega_k (t-t_0)} - 1)/\Omega_k$,
where $\Omega_k \equiv \omega_k - \omega_\ell$.
Thus, by using Eqs.~(\ref{KwHd}) and (\ref{coherent_state}), we rewrite Eq.~(\ref{work_flow_2}) as
\begin{align}
J_{\rm d \leftarrow c}^{\rm work}
&= -2 \hbar \omega_\ell \sum_k {\rm Im} \bigl[ g_k \langle \hat{c}^\dag \rangle_{\rm c}^{\rm ss} {\rm Tr_d} \{ \rho_{\rm d}(t)\hat{d}_k \} \bigr]
\notag\\
&\simeq -2 \hbar \omega_\ell \sum_k
{\rm Im} \bigl[ g_k \langle \hat{c}^\dag \rangle_{\rm c}^{\rm ss} \langle \gamma_k(t) | \hat{d}_k | \gamma_k(t) \rangle \bigr]
\notag\\
&= 2 \hbar \omega_\ell \big| \langle \hat{c} \rangle_{\rm c}^{\rm ss} \big|^2 \sum_k | g_k |^2 \frac{\sin \Omega_k (t-t_0)}{\Omega_k}
\notag\\
&= \frac{\hbar \omega_\ell \big| \langle \hat{c} \rangle_{\rm c}^{\rm ss} \big|^2}{\pi}
\int_{-\infty}^\infty d\omega \Gamma_{\rm d}(\omega) \frac{\sin \Omega (t-t_0)}{\Omega}.
\end{align}
In the last line, we have defined $\Omega \equiv \omega - \omega_\ell$
and used the definition of $\Gamma_{\rm d}(\omega)$ [Eq.~(\ref{spectral_drain})].

Now we take the wide-band limit: $\Gamma_{\rm d}(\omega) = \kappa$.
Then we finally obtain the desired result:
\begin{align}
J_{\rm d \leftarrow c}^{\rm work}
&= \frac{\kappa \hbar \omega_\ell \big| \langle \hat{c} \rangle_{\rm c}^{\rm ss} \big|^2}{\pi}
\int_{-\infty}^\infty d\Omega \frac{\sin \Omega (t-t_0)}{\Omega}
\notag\\
&= \kappa \hbar \omega_\ell \big| \langle \hat{c} \rangle_{\rm c}^{\rm ss} \big|^2,
\end{align}
where we have used $\int_{-\infty}^\infty d\Omega (1/\Omega) \sin \Omega (t-t_0) = \pi$ for $t>t_0$.

\subsection{Derivation of Eq.~(\ref{heat_flow_explicit})}

Next we derive the explicit form of the heat flow $J_{\rm d \leftarrow c}^{\rm heat}$.
In the definition of $J_{\rm d \leftarrow c}^{\rm heat}$, Eq.~(\ref{heat_flow}),
since $\mathcal{K}_{\rm h}$ is on the second order in the coupling,
it is sufficient to take only $H_{\rm d}$ (neglect $\bar{H}_\text{c-d}$) and  evaluate $\rho_{\rm d}(t)$ on the zeroth order.
On the zeroth order we have  $\rho_{\rm d}(t) \simeq e^{(H_{\rm d} - \hbar \omega_\ell \hat{N}_{\rm d}) (t-t_0)/i\hbar} | {\rm vac} \rangle \langle {\rm vac} | e^{-(H_{\rm d} - \hbar \omega_\ell \hat{N}_{\rm d}) (t-t_0)/i\hbar} = | {\rm vac} \rangle \langle {\rm vac} |$.
We thus rewrite $J_{\rm d \leftarrow c}^{\rm heat}$ on the second order as
\begin{align}
J_{\rm d \leftarrow c}^{\rm heat} \simeq {\rm Tr_d} \bigl\{ H_{\rm d} \mathcal{K}_{\rm h}(t) | {\rm vac} \rangle \langle {\rm vac} | \bigr\}.
\label{J_heat_2nd}
\end{align}

To proceed further, we note that $\mathcal{K}_{\rm h}(t)$ acts on any drain operator $\hat{O}_{\rm d}$ as
\begin{widetext}
\begin{align}
\mathcal{K}_{\rm h}(t) \hat{O}_{\rm d}
&= -\sum_{k,k'} \int_0^{t-t_0} d\tau {\rm Tr_c} \Bigl[ g_k \delta\Check{c}^\dag(\tau) \hat{d}_k + g_k^* \delta\Check{c}(\tau) \hat{d}_k^\dag ,
\bigl[ g_{k'} \delta\hat{c}^\dag \hat{d}_{k'} e^{i\Omega_{k'}\tau} + g_{k'}^* \delta\hat{c} \hat{d}_{k'}^\dag e^{-i\Omega_{k'}\tau},
\rho_{\rm c}^{\rm ss} \otimes \hat{O}_{\rm d} \bigr] \Bigr],
\end{align}
where $\Omega_k \equiv \omega_k - \omega_\ell$.
Substituting $H_d = \sum_k \hbar \omega_k \hat{d}_k^\dag \hat{d}_k$ and this equation into Eq.~(\ref{J_heat_2nd}), we obtain
\begin{align}
J_{\rm d \leftarrow c}^{\rm heat}
=& \int_0^{t-t_0} d\tau \sum_k |g_k|^2 \hbar \omega_k
\bigl( \langle \delta\hat{c}^\dag \delta\Check{c}(\tau) \rangle_{\rm c}^{\rm ss} e^{i\Omega_k \tau}
+ \langle \delta\Check{c}^\dag(\tau) \delta\hat{c} \rangle_{\rm c}^{\rm ss} e^{-i\Omega_k \tau} \bigr)
\notag\\
=& \frac{1}{2\pi} \int_0^{t-t_0} d\tau \int_{-\infty}^\infty d\omega \Gamma_{\rm d}(\omega) \hbar \omega
\bigl( \langle \delta\hat{c}^\dag \delta\Check{c}(\tau) \rangle_{\rm c}^{\rm ss} e^{i\Omega \tau}
+ \langle \delta\Check{c}^\dag(\tau) \delta\hat{c} \rangle_{\rm c}^{\rm ss} e^{-i\Omega \tau} \bigr)
\notag\\
=& \frac{1}{2\pi} \int_0^{t-t_0} d\tau \int_{-\infty}^\infty d\omega \Gamma_{\rm d}(\omega) \hbar (\omega_\ell + \Omega)
\bigl( \langle \delta\hat{c}^\dag \delta\Check{c}(\tau) \rangle_{\rm c}^{\rm ss} e^{i\Omega \tau}
+ \langle \delta\Check{c}^\dag(\tau) \delta\hat{c} \rangle_{\rm c}^{\rm ss} e^{-i\Omega \tau} \bigr)
\notag\\
=& \frac{\hbar \omega_\ell}{2\pi} \int_0^{t-t_0} d\tau \int_{-\infty}^\infty d\omega \Gamma_{\rm d}(\omega)
\bigl( \langle \delta\hat{c}^\dag \delta\Check{c}(\tau) \rangle_{\rm c}^{\rm ss} e^{i\Omega \tau}
+ \langle \delta\Check{c}^\dag(\tau) \delta\hat{c} \rangle_{\rm c}^{\rm ss} e^{-i\Omega \tau} \bigr)
\notag\\
&+ \frac{1}{2\pi} \int_{-\infty}^\infty d\omega \Gamma_{\rm d}(\omega) i \hbar
\Bigl[ -\langle \delta\hat{c}^\dag \delta\Check{c}(\tau) \rangle_{\rm c}^{\rm ss} e^{i\Omega \tau}
+ \langle \delta\Check{c}^\dag(\tau) \delta\hat{c} \rangle_{\rm c}^{\rm ss} e^{-i\Omega \tau} \Bigr]_{\tau = 0}^{t-t_0}
\notag\\
&+ \frac{1}{2\pi} \int_0^{t-t_0} d\tau \int_{-\infty}^\infty d\omega \Gamma_{\rm d}(\omega) i \hbar
\Bigl( e^{i\Omega \tau} \frac{d}{d\tau} \langle \delta\hat{c}^\dag \delta\Check{c}(\tau) \rangle_{\rm c}^{\rm ss}
- e^{-i\Omega \tau} \frac{d}{d\tau} \langle \delta\Check{c}^\dag(\tau) \delta\hat{c} \rangle_{\rm c}^{\rm ss} \Bigr),
\end{align}
where we have used the definition of $\Gamma_{\rm d}(\omega)$ [Eq.~(\ref{spectral_drain})] and defined $\Omega \equiv \omega - \omega_\ell$.

Now we take the wide-band limit: $\Gamma_{\rm d}(\omega) = \kappa$.
Then we obtain
\begin{align}
J_{\rm d \leftarrow c}^{\rm heat}
=& \frac{\hbar \omega_\ell \kappa}{2\pi} \int_0^{t-t_0} d\tau \int_{-\infty}^\infty d\Omega
\bigl( \langle \delta\hat{c}^\dag \delta\Check{c}(\tau) \rangle_{\rm c}^{\rm ss} e^{i\Omega \tau}
+ \langle \delta\Check{c}^\dag(\tau) \delta\hat{c} \rangle_{\rm c}^{\rm ss} e^{-i\Omega \tau} \bigr)
\notag\\
&+ \frac{i \hbar \kappa}{2\pi} \int_{-\infty}^\infty d\Omega
\bigl( -\langle \delta\hat{c}^\dag \delta\Check{c}(\tau) \rangle_{\rm c}^{\rm ss} e^{i\Omega (t-t_0)}
+ \langle \delta\Check{c}^\dag(\tau) \delta\hat{c} \rangle_{\rm c}^{\rm ss} e^{-i\Omega (t-t_0)} \bigr)
\notag\\
&+ \frac{i \hbar \kappa}{2\pi} \int_0^{t-t_0} d\tau \int_{-\infty}^\infty d\Omega
\Bigl( e^{i\Omega \tau} \frac{d}{d\tau} \langle \delta\hat{c}^\dag \delta\Check{c}(\tau) \rangle_{\rm c}^{\rm ss}
- e^{-i\Omega \tau} \frac{d}{d\tau} \langle \delta\Check{c}^\dag(\tau) \delta\hat{c} \rangle_{\rm c}^{\rm ss} \Bigr)
\notag\\
=& \frac{\hbar \omega_\ell \kappa}{2\pi} \int_0^{t-t_0} d\tau
\bigl( \langle \delta\hat{c}^\dag \delta\Check{c}(\tau) \rangle_{\rm c}^{\rm ss} \delta(\tau)
+ \langle \delta\Check{c}^\dag(\tau) \delta\hat{c} \rangle_{\rm c}^{\rm ss} \delta(\tau) \bigr)
\notag\\
&+ \frac{i \hbar \kappa}{2\pi} \int_0^{t-t_0} d\tau
\Bigl( \delta(\tau) \frac{d}{d\tau} \langle \delta\hat{c}^\dag \delta\Check{c}(\tau) \rangle_{\rm c}^{\rm ss}
- \delta(\tau) \frac{d}{d\tau} \langle \delta\Check{c}^\dag(\tau) \delta\hat{c} \rangle_{\rm c}^{\rm ss} \Bigr)
\notag\\
=& ~\hbar \omega_\ell \kappa \bigl\langle \delta\hat{c}^\dag \delta\hat{c} \bigr\rangle_{\rm c}^{\rm ss}
+ \hbar \kappa \frac{d}{d\tau} {\rm Im} \bigl\langle \delta\Check{c}^\dag(\tau) \delta\hat{c} \bigr\rangle_{\rm c}^{\rm ss} \Big|_{\tau=0},
\end{align}
where we have used $\int_{-\infty}^\infty d\Omega e^{\pm i\Omega (t-t_0)} = 0$ for $t>t_0$ in the second equality
and $\int_0^{t-t_0} d\tau \delta(\tau) f(\tau) = (1/2) f(0)$ in the third equality.
Finally by noting
${\rm Im} \langle \delta\Check{c}^\dag(\tau) \delta\hat{c} \rangle_{\rm c}^{\rm ss}
= {\rm Im} \bigl( \langle \Check{c}^\dag(\tau) \hat{c} \rangle_{\rm c}^{\rm ss} - | \langle \hat{c} \rangle_{\rm c}^{\rm ss} |^2 \bigr)
= {\rm Im} \langle \Check{c}^\dag(\tau) \hat{c} \rangle_{\rm c}^{\rm ss}$
(because $| \langle \hat{c} \rangle_{\rm c}^{\rm ss} |^2$ is a real number),
we obtain the desired result, Eq.~(\ref{heat_flow_explicit}).
\end{widetext}

\section{Derivation of TCL quantum master equation~(\ref{TCLQME_drain}) for drain}
\label{derivation_TCLQME}

For completeness, we here show a detailed derivation of the TCL-QME~(\ref{TCLQME_drain}) for the drain.
It is almost the same as the standard derivation \cite{BreuerPetruccione}.
Main differences are: (i) the starting-point equation is not a von Neumann equation but the QME~(\ref{cQED+drain}) for the composite system,
and (ii) the coupling between the cQED and drain does not vanish when taking its average in the environment (cQED) steady state.

To perform a perturbation expansion later, we express the coupling strength between the cQED and drain by $\epsilon$;
i.e., we rewrite the interaction Hamiltonian as $H_\text{c-d} \to \epsilon H_\text{c-d}$.
Then we may split the Liouvillian $\mathcal{M}$ into terms on the order of $\epsilon$ as
$\mathcal{M} = \mathcal{M}_0 + \epsilon \mathcal{M}_1 + \epsilon^2 \mathcal{M}_2$,
where $\mathcal{M}_0 = \mathcal{L} + \mathcal{K}_0$, $\mathcal{M}_1 = \mathcal{M}_{\rm int}$, and $\mathcal{M}_2 = - \mathcal{L}_{\rm d}$.

\subsection{TCL equation}

We first drive a TCL equation which is valid up to the all order in $\epsilon$.
We transform to an ``interaction picture'' by
\begin{align}
\breve{\rho}_{\rm c+d}(t) = e^{-\mathcal{M}_0 t} \rho_{\rm c+d}(t).
\end{align}
In this picture the starting-point QME~(\ref{cQED+drain}) becomes
\begin{align}
\frac{d}{dt} \breve{\rho}_{\rm c+d}(t) &= \breve{\mathcal{M}}_{\rm p}(t) \breve{\rho}_{\rm c+d}(t),
\label{interaction_pic}
\end{align}
where $\breve{\mathcal{M}}_{\rm p}(t) = \epsilon \breve{\mathcal{M}}_1(t) + \epsilon^2 \breve{\mathcal{M}}_2(t)$,
and $\breve{\mathcal{M}}_j(t) = e^{-\mathcal{M}_0 t} \mathcal{M}_j e^{\mathcal{M}_0 t}$ ($j = 1, 2$).
The formal backward solution of this equation is
\begin{align}
\breve{\rho}_{\rm c+d}(t') &= \mathcal{B}(t', t) \breve{\rho}_{\rm c+d}(t),
\label{backward_solution}
\end{align}
where $\mathcal{B}(t', t) = {\rm T}_\rightarrow \exp \int_{t'}^t dt'' \breve{\mathcal{M}}_{\rm p}(t'')$ for $t>t'$,
and ${\rm T}_\rightarrow$ is the antichronological time ordering.

We here introduce projection superoperators $\mathcal{P}$ and $\mathcal{Q} = 1 - \mathcal{P}$.
(We will specify the explicit form of these superoperators later.)
Then we may split Eq.~(\ref{interaction_pic}) into two parts:
\begin{align}
\frac{d}{dt} \mathcal{P} \breve{\rho}_{\rm c+d}(t)
&= \mathcal{P} \breve{\mathcal{M}}_{\rm p}(t) \mathcal{P} \breve{\rho}_{\rm c+d}(t)
+ \mathcal{P} \breve{\mathcal{M}}_{\rm p}(t) \mathcal{Q} \breve{\rho}_{\rm c+d}(t),
\label{Prho}
\\
\frac{d}{dt} \mathcal{Q} \breve{\rho}_{\rm c+d}(t)
&= \mathcal{Q} \breve{\mathcal{M}}_{\rm p}(t) \mathcal{P} \breve{\rho}_{\rm c+d}(t)
+ \mathcal{Q} \breve{\mathcal{M}}_{\rm p}(t) \mathcal{Q} \breve{\rho}_{\rm c+d}(t).
\label{Qrho}
\end{align}
The formal solution of the latter equation is
\begin{align}
\mathcal{Q} \breve{\rho}_{\rm c+d}(t) = &~ \mathcal{G}(t, t_0) \mathcal{Q} \breve{\rho}_{\rm c+d}(t_0)
\notag\\
&~+ \int_{t_0}^t dt' \mathcal{G}(t, t') \mathcal{Q} \breve{\mathcal{M}}_{\rm p}(t') \mathcal{P} \breve{\rho}_{\rm c+d}(t'),
%\label{Qrho_formal}
\notag
\end{align}
where $\mathcal{G}(t, t') = {\rm T}_\leftarrow \exp \int_{t'}^t dt'' \mathcal{Q} \breve{\mathcal{M}}_{\rm p}(t'')$ for $t>t'$,
and ${\rm T}_\leftarrow$ is the chronological time ordering.
By substituting Eq.~(\ref{backward_solution}) into this solution and by doing simple algebraic calculation, we obtain
\begin{align}
\bigl[1 - \Sigma(t) \bigr] \mathcal{Q} \breve{\rho}_{\rm c+d}(t)
= \Sigma(t) \mathcal{P} \breve{\rho}_{\rm c+d}(t) + \mathcal{G}(t, t_0) \mathcal{Q} \breve{\rho}_{\rm c+d}(t_0),
\notag
\end{align}
where $\Sigma(t) \equiv \int_{t_0}^t dt' \mathcal{G}(t, t') \mathcal{Q} \breve{\mathcal{M}}_{\rm p}(t') \mathcal{P} \mathcal{B}(t', t)$.

In the conditions of the weak coupling ($\epsilon \ll 1$) and/or short time ($t \simeq t_0$),
$\Sigma(t)$ is very small, so that it is reasonable to assume the existence of the inverse of $1 - \Sigma(t)$.
Then the above equation becomes
\begin{align}
\mathcal{Q} \breve{\rho}_{\rm c+d}(t)
=&~ \bigl[1 - \Sigma(t) \bigr]^{-1} \Sigma(t) \mathcal{P} \breve{\rho}_{\rm c+d}(t)
\notag\\
&+ \bigl[1 - \Sigma(t) \bigr]^{-1} \mathcal{G}(t, t_0) \mathcal{Q} \breve{\rho}_{\rm c+d}(t_0).
\notag
\end{align}
Substituting this equation into Eq.~(\ref{Prho}), we obtain the TCL equation:
\begin{align}
\frac{d}{dt} \mathcal{P} \breve{\rho}_{\rm c+d}(t)
=&~ \mathcal{P} \breve{\mathcal{M}}_{\rm p}(t) \bigl[1 - \Sigma(t) \bigr]^{-1} \mathcal{P} \breve{\rho}_{\rm c+d}(t)
\notag\\
&+ \mathcal{P} \breve{\mathcal{M}}_{\rm p}(t) \bigl[1 - \Sigma(t) \bigr]^{-1} \mathcal{G}(t, t_0) \mathcal{Q} \breve{\rho}_{\rm c+d}(t_0).
\label{TCL_eq}
\end{align}

\subsection{Second-order perturbation expansion}

We next evaluate $\bigl[1 - \Sigma(t) \bigr]^{-1}$ in Eq.~(\ref{TCL_eq}) by the perturbation theory in terms of $\epsilon$.
Since the lowest order term in the $\epsilon$-expansion of $\Sigma(t)$ is
$\epsilon \Sigma_1(t) \equiv \epsilon \int_{t_0}^t dt' \mathcal{Q} \breve{\mathcal{M}}_1(t') \mathcal{P}$,
we have $\bigl[1-\Sigma(t) \bigr]^{-1} = 1 + \epsilon \Sigma_1(t) + O(\epsilon^2)$.
Then, substituting this into the first term of the right-hand side of Eq.~(\ref{TCL_eq}), we obtain in $O(\epsilon^2)$
\begin{align}
\frac{d}{dt} \mathcal{P} \breve{\rho}_{\rm c+d}(t)
=&~ \epsilon \mathcal{P} \breve{\mathcal{M}}_1(t) \mathcal{P} \breve{\rho}_{\rm c+d}(t)
+ \epsilon^2 \mathcal{P} \breve{\mathcal{M}}_2(t) \mathcal{P} \breve{\rho}_{\rm c+d}(t)
\notag\\
&+ \epsilon^2 \int_{t_0}^t dt' \mathcal{P} \breve{\mathcal{M}}_1(t) \mathcal{Q} \breve{\mathcal{M}}_1(t') \mathcal{P} \breve{\rho}_{\rm c+d}(t)
\notag\\
&+ \mathcal{P} \breve{\mathcal{M}}_{\rm p}(t) \bigl[1-\Sigma(t) \bigr]^{-1} \mathcal{G}(t, t_0) \mathcal{Q} \breve{\rho}_{\rm c+d}(t_0).
\label{TCL_eq_2nd}
\end{align}

\subsection{Projection onto the steady state of the cQED}

We now define the explicit form of the projection superoperator $\mathcal{P}$,
such that the following equation holds for any operator $\hat{O}_{\rm c+d}$:
\begin{align}
\mathcal{P} \hat{O}_{\rm c+d} = \rho_{\rm c}^{\rm ss} \otimes {\rm Tr_c} \hat{O}_{\rm c+d}.
\end{align}

Furthermore, we assume that the initial state of the composite system is given by
$\breve{\rho}_{\rm c+d}(t_0) = \rho_{\rm sys}^{\rm ss} \otimes \breve{\rho}_{\rm d}(t_0)$.
In this case, $\mathcal{Q} \breve{\rho}_{\rm c+d}(t_0) = 0$ holds,
so that the last term of Eq.~(\ref{TCL_eq_2nd}) vanishes.

In the below we investigate the following three terms, which appear in Eq.~(\ref{TCL_eq_2nd}):
$\mathcal{P} \breve{\mathcal{M}}_1(t) \mathcal{P}$, $\mathcal{P} \breve{\mathcal{M}}_2(t) \mathcal{P}$,
and $\mathcal{P} \breve{\mathcal{M}}_1(t) \mathcal{Q} \breve{\mathcal{M}}_1(t') \mathcal{P}$.

First we calculate $\mathcal{P} \breve{\mathcal{M}}_2(t)$:
\begin{align}
&\mathcal{P} \breve{\mathcal{M}}_2(t) \hat{O}_{\rm c+d}
\notag\\
&= - \rho_{\rm c}^{\rm ss} \otimes e^{-\mathcal{K}_0 (t-t_0)} {\rm Tr_c} \bigl\{ e^{-\mathcal{L} (t-t_0)} \mathcal{L}_{\rm d} e^{\mathcal{M}_0 (t-t_0)} \hat{O}_{\rm c+d} \bigr\}
\notag\\
&= 0,
\label{PM2}
\end{align}
where we have used $e^{-\mathcal{M}_0 (t-t_0)} = e^{-\mathcal{L} (t-t_0)} e^{-\mathcal{K}_0 (t-t_0)}$,
the trace preserving property of $e^{-\mathcal{L} (t-t_0)}$, and ${\rm Tr_c} \{ \mathcal{L}_{\rm d} \hat{O}_{\rm c} \} = 0$.
Equation (\ref{PM2}) implies $\mathcal{P} \breve{\mathcal{M}}_2(t) \mathcal{P} = 0$.

Next we calculate $\mathcal{P} \breve{\mathcal{M}}_1(t) \mathcal{P}$:
\begin{align}
&\mathcal{P} \breve{\mathcal{M}}_1(t) \mathcal{P} \hat{O}_{\rm c+d}
\notag\\
&= \rho_{\rm c}^{\rm ss} \otimes e^{-\mathcal{K}_0 (t-t_0)} {\rm Tr_c} \left\{ \frac{1}{i\hbar} [ H_\text{c-d} ,
e^{\mathcal{M}_0 (t-t_0)} \mathcal{P} \hat{O}_{\rm c+d} ] \right\}
\notag\\
&= \rho_{\rm c}^{\rm ss} \otimes e^{-\mathcal{K}_0 (t-t_0)} {\rm Tr_c} \left\{ \frac{1}{i\hbar} [ H_\text{c-d} ,
\rho_{\rm c}^{\rm ss} \otimes e^{\mathcal{K}_0 (t-t_0)} {\rm Tr_c} \hat{O}_{\rm c+d} ] \right\}
\notag\\
&= \rho_{\rm c}^{\rm ss} \otimes e^{-\mathcal{K}_0 (t-t_0)} \frac{1}{i\hbar} [ \bar{H}_\text{c-d} ,
e^{\mathcal{K}_0 (t-t_0)} {\rm Tr_c} \hat{O}_{\rm c+d} ]
\notag\\
&= \rho_{\rm c}^{\rm ss} \otimes e^{-\mathcal{K}_0 (t-t_0)} \bar{\mathcal{M}}_1 e^{\mathcal{K}_0 (t-t_0)} {\rm Tr_c} \hat{O}_{\rm c+d},
\label{PM1P}
\end{align}
where we have used $e^{\mathcal{L} (t-t_0)} \rho_{\rm c}^{\rm ss} =\rho_{\rm c}^{\rm ss}$
and defined $\bar{\mathcal{M}}_1$ by $\bar{\mathcal{M}}_1 \hat{O}_{\rm d} = (1/i\hbar) [ \bar{H}_\text{c-d}, \hat{O}_{\rm d} ]$.

Finally we calculate $\mathcal{P} \breve{\mathcal{M}}_1(t) \mathcal{Q} \breve{\mathcal{M}}_1(t') \mathcal{P}$.
By noting that Eq.~(\ref{PM1P}) implies $\mathcal{P} \breve{\mathcal{M}}_1(t) \mathcal{P} = \mathcal{P} \bar{\mathcal{M}}_1(t)
= \bar{\mathcal{M}}_1(t) \mathcal{P}$,
we obtain $\mathcal{P} \breve{\mathcal{M}}_1(t) \mathcal{Q} = \mathcal{P} \breve{\mathcal{M}}'_1(t)$
and $\mathcal{Q} \breve{\mathcal{M}}_1(t) \mathcal{P} = \breve{\mathcal{M}}'_1(t) \mathcal{P}$,
where $\bar{\mathcal{M}}_1(t) \equiv e^{-\mathcal{M}_0 (t-t_0)} \bar{\mathcal{M}}_1 e^{\mathcal{M}_0 (t-t_0)}$,
$\breve{\mathcal{M}}'_1(t) \equiv e^{-\mathcal{M}_0 (t-t_0)} \breve{\mathcal{M}}'_1 e^{\mathcal{M}_0 (t-t_0)}$,
and $\breve{\mathcal{M}}'_1 \equiv \breve{\mathcal{M}}_1 - \bar{\mathcal{M}}_1$.
These equations lead to
\begin{align}
\mathcal{P} \breve{\mathcal{M}}_1(t) \mathcal{Q} \breve{\mathcal{M}}_1(t') \mathcal{P}
&= \mathcal{P} \breve{\mathcal{M}}'_1(t) \breve{\mathcal{M}}'_1(t') \mathcal{P}.
\end{align}

\subsection{TCL quantum master equation}

By applying the results in the previous subsection to Eq.~(\ref{TCL_eq_2nd}), we obtain
\begin{widetext}
\begin{align}
\frac{d}{dt} \mathcal{P} \breve{\rho}(t)
= \epsilon \rho_{\rm c}^{\rm ss} \otimes e^{-\mathcal{K}_0 (t-t_0)} \bar{\mathcal{M}}_1 e^{\mathcal{K}_0 (t-t_0)} \breve{\rho}_{\rm d}(t)
+ \epsilon^2 \rho_{\rm c}^{\rm ss} \otimes {\rm Tr_c} \int_{t_0}^t dt' e^{-\mathcal{M}_0 (t-t_0)} \mathcal{M}'_1 e^{\mathcal{M}_0 (t-t')} \mathcal{M}'_1
e^{\mathcal{M}_0 (t'-t_0)} [ \rho_{\rm c}^{\rm ss} \otimes \breve{\rho}_{\rm d}(t) ],
\end{align}
where $\breve{\rho}_{\rm d}(t) \equiv {\rm Tr_c} \breve{\rho}_{\rm c+d}(t)$.
We take ${\rm Tr_c}$ of the above equation to obtain the second-order TCL-QME in the interaction picture:
\begin{align}
\frac{d}{dt} \breve{\rho}_{\rm d}(t)
= \epsilon e^{-\mathcal{K}_0 (t-t_0)} \bar{\mathcal{M}}_1 e^{\mathcal{K}_0 (t-t_0)} \breve{\rho}_{\rm d}(t)
+ \epsilon^2 {\rm Tr_c} \int_{t_0}^t dt' e^{-\mathcal{K}_0 (t-t_0)} \mathcal{M}'_1 e^{\mathcal{M}_0 (t-t')} \mathcal{M}'_1
\bigl[ \rho_{\rm c}^{\rm ss} \otimes e^{\mathcal{K}_0 (t'-t_0)} \breve{\rho}_{\rm d}(t) \bigr].
\end{align}
Finally, by going back to the Sch\"odinger picture, we eventually obtain the desired result, Eq.~(\ref{TCLQME_drain}).
\end{widetext}


\begin{thebibliography}{99}

\bibitem{Fermi}
E. Fermi, ``Thermodynamics'' (Dover, New York, 1956).

\bibitem{Callen}
H. B. Callen, ``Thermodynamics and an Introduction to Thermostatistics,'' 2nd ed. (Wiley, New York, 1985).

\bibitem{GyftopoulosBeretta}
E. P. Gyftopoulos and G. P. Beretta,``Thermodynamics: Foundations and Applications'' (Dover, New York, 2005).


\bibitem{Clausius}
R. Clausius,
``The Mechanical Theory of Heat
with its Application to the Steam Engine and to the Physical Properties of Bodies'' (London, 1867).

\bibitem{Thomson}
W. Thomson,
``Mathematical and Physical papers,'' vol. 1 (Cambridge University Press, Cambridge, 1882).

\bibitem{Einstein}
A. Einstein, Annalen der Physik. {\bf 332}, 132 (1905).


\bibitem{Kosloff}
R. Kosloff, Entropy {\bf 15}, 2100 (2013).

\bibitem{Mahler}
G. Mahler, ``Quantum Thermodynamic Processes: Energy and Information Flow at the Nanoscale'' (Pan Stanford, Singapore, 2015).

\bibitem{VinjanampathyAnders}
S. Vinjanampathy and J. Anders, Contemporary Physics, {\bf 57},  1 (2016).

\bibitem{Goold_etal}
J. Goold, M. Huber, A. Riera, L. del Rio, P. Skrzypczyk,
J. Phys. A {\bf 49}, 143001 (2016).


\bibitem{vonNeumann}
J. von Neumann,
``Mathematical Foundations of Quantum Mechanics'' (Princeton University Press, New Jersey, 1955)


\bibitem{Araki}
H. Araki, Commun. math. Phys. {\bf 38}, 1 (1974).

\bibitem{HatsopoulosGyftopoulos}
G. N. Hatsopoulos and E. P. Gyftopoulos, Found. Phys., {\bf 6}, 15 (1976);
{\it ibid} {\bf 6}, 127 (1976); {\bf 6}, 439 (1976); {\bf 6}, 561 (1976).

\bibitem{PuszWoronowicz}
W. Pusz and S. L. Woronowicz, Commun. math. Phys. {\bf 58}, 273 (1978).

\bibitem{Lenard}
A. Lenard, J. Stat. Phys., {\bf 19}, 575 (1978).


\bibitem{Alicki1976}
R. Alicki, Rep. Math. Phys. {\bf 10}, 249 (1976).

\bibitem{Spohn}
H. Spohn, J. Math. Phys. {\bf 19} 1227 (1978).


\bibitem{SpohnLebowitz}
H. Spohn and J. L. Lebowitz, Adv. Chem. Phys. {\bf 38}, 109 (1978).

\bibitem{Alicki1979}
R. Alicki, J. Phys. A {\bf 12}, L103 (1979).

\bibitem{Kosloff1984}
R. Kosloff, J. Chem. Phys. {\bf 80}, 1625 (1984).


\bibitem{ScovilSchulz-DuBois}
H. E. D. Scovil and E. O. Schulz-DuBois, Phys. Rev. Lett. {\bf 2}, 262 (1959).

\bibitem{Geusic_etal}
J. E. Geusic, E. O. Schulz-DuBois, and H. E. D. Scovil, Phys. Rev. {\bf 156}, 343 (1967).

\bibitem{Rmasey}
N. F. Ramsey, Phys. Rev. {\bf 103}, 20 (1956).


\bibitem{AllahverdyanNieuwenhuizen}
A. E. Allahverdyan and Th. M. Nieuwenhuizen, Phys. Rev. Lett. {\bf 85}, 1799 (2000).

\bibitem{Gemmer_etal}
J. Gemmer, A. Otte, and G. Mahler, Phys. Rev. Lett. {\bf 86}, 1927 (2001).

\bibitem{Horodecki_etal}
M. Horodecki, J. Oppenheim, and R. Horodecki, Phys. Rev. Lett. {\bf 89}, 240403 (2002).

\bibitem{Scully_etal2003}
M. O. Scully, M. S. Zubairy, G. S. Agarwal, and H. Walther, Science {\bf 299}, 862 (2003).

\bibitem{Allahverdyan_etal}
A. E. Allahverdyan, R. Balian, and Th. M. Nieuwenhuizen, Europhys. Lett., {\bf 67}, 565 (2004).

\bibitem{Weimer_etal}
H. Weimer, M. J. Henrich, F. Rempp, H. Schr\"oder, and G. Mahler, Europhys. Lett., {\bf 83}, 30008 (2008).

\bibitem{Scully}
M. O. Scully, Phys. Rev. Lett. {\bf 104}, 207701 (2010).

\bibitem{Scully_etal2011}
M. O. Scully, K. R. Chapin, K. E. Dorfman, M. B. Kim, and A. Svidzinsky, Proc. Natl. Acad. Sci. U.S.A. {\bf 108}, 15097 (2011).
%[L]

\bibitem{Beretta}
G. P. Beretta, Europhys. Lett., {\bf 99}, 20005 (2012).

\bibitem{Harbola_etal}
U. Harbola, S. Rahav, and S. Mukamel, Europhys. Lett. {\bf 99}, 50005 (2012).

\bibitem{Dorfman_etal}
K. E. Dorfman, D. V. Voronine, S. Mukamel, and M. O. Scully, Proc. Natl. Acad. Sci. U.S.A. {\bf 110}, 2746 (2013).
%[L]

\bibitem{Gelbwaser-Klimovsky_etal}
D. Gelbwaser-Klimovsky, R. Alicki, and G. Kurizki, Europhys. Lett. {\bf 103}, 60005 (2013).

\bibitem{GoswamiHarbola}
H. P. Goswami and U. Harbola, Phys. Rev. A {\bf 88}, 013842 (2013).
%[L]

\bibitem{Aberg}
J. \r{A}berg, Phys. Rev. Lett. {\bf 113}, 150402 (2014).

\bibitem{Correa_etal}
L. A. Correa, J. P. Palao, D. Alonso, and G. Adesso, Sci. Rep. {\bf 4}, 3949 (2014).

\bibitem{Rosnagel_etal}
J. Ro{\ss}nagel, O. Abah, F. Schmidt-Kaler, K. Singer, and E. Lutz, Phys. Rev. Lett. {\bf 112}, 030602 (2014).

\bibitem{Mitchison_etal}
M. T. Mitchison, M. P. Woods, J. Prior, and M. Huber, New J. Phys. {\bf 17}, 115013 (2015).

\bibitem{BraskBrunner}
J. B. Brask and N. Brunner, Phys. Rev. E {\bf 92}, 062101 (2015).

\bibitem{Uzdin_etal}
R. Uzdin, A. Levy, and R. Kosloff, Phys. Rev. X {\bf 5}, 031044 (2015).

\bibitem{Korzekw_etal}
K. Korzekwa, M. Lostaglio, J. Oppenheim, and D. Jennings, New J. Phys. {\bf 18}, 023045 (2016).

\bibitem{Dag_etal}
C. B. Dag, W. Niedenzu, O. E. Mustecaplioglu, and G. Kurizki,
Entropy, {\bf 18}, 244 (2016).


\bibitem{BreuerPetruccione}
H. P. Breuer and F. Petruccione, ``The Theory of Open Quantum Systems''
(Oxford University Press, Oxford, 2002).


% condition for relaxing
\bibitem{Frigerio}
A. Frigerio, Lett. Math. Phys., {\rm 2}, 79 (1977).

\bibitem{RivasHuelga}
A. Rivas and S. F. Huelga, ``Open Quantum Systems: An Introduction''
(Springer, Heidelberg Dordrecht London New York, 2012)

\bibitem{NakataniOgawa}
M. Nakatani and T. Ogawa, J. Phys. Soc. Jpn. {\bf 79}, 084401 (2010).


\bibitem{Davidson}
N. R. Davidson, ``Statistical Mechanics'' (McGraw-Hill, New York, 1962).


\bibitem{Sekimoto}
K. Sekimoto, ``Stochastic Energetics'' (Springer-Verlag Berlin Heidelberg, 2010).

\bibitem{BustamanteLiphardtRitort}
C. Bustamante, J. Liphardt, and F. Ritort, Physics Today, {\bf 58}, 43 (2005).


% singular coupling limit
\bibitem{GoriniKossakowski}
V. Gorini and A. Kossakowski, J. Math. Phys. {\bf 17}, 1298 (1976).

% singular coupling limit
\bibitem{FrigerioGorini}
A. Frigerio and V. Gorini, J. Math. Phys. {\bf 17}, 2123 (1976).


% hierarchy equation
\bibitem{Tanimura}
Y. Tanimura, J. Phys. Soc. Jpn. {\bf 75}, 082001 (2006).

% hierarchy equation
\bibitem{KatoTanimura}
A. Kato and Y. Tanimura, arXiv:1609.08783.

% Gibbs reservoir
\bibitem{HorowitzEsposito}
J. M. Horowitz and M. Esposito, Phys Rev E {\bf 94}, 020102 (2016).

\bibitem{BerettaGyftopoulos}
G. P. Beretta and E. P. Gyftopoulos, Journal of Energy Resources Technology, {\bf 137}, 021005 (2015).



\end{thebibliography}
\end{document}